\documentclass[preprint,times]{elsarticle}
\usepackage{hyperref}
\usepackage{amsmath}
\journal{Journal of Computational Physics}
\biboptions{numbers,sort&compress}

\def\xbar{\mathbf{\bar{x}}}
\def\xbf{\mbox{$\mathbf{x}$}}
\def\gammabf{\mbox{$\boldsymbol{\gamma}$}}
\def\kbf{\mbox{$\mathbf{k}$}}
\def\sbf{\mbox{$\mathbf{s}$}}
\def\ubf{\mbox{$\mathbf{u}$}}
\def\Dbf{\mbox{$\mathbf{D}$}}
\def\Mbf{\mbox{$\mathbf{M}$}}
\def\Fbf{\mbox{$\mathbf{F}$}}
\def\Jbf{\mbox{$\mathbf{J}$}}
\def\Abf{\mbox{$\mathbf{A}$}}

\DeclareMathOperator{\diag}{diag}
\DeclareMathOperator{\Tr}{Tr}

\numberwithin{equation}{section}

\usepackage[dvipsnames]{xcolor}

\begin{document}

\begin{frontmatter}

\title{Formulation of discontinuous Galerkin methods for relativistic
astrophysics}

\author[cornell,caltech]{Saul A. Teukolsky}
\ead{saul@astro.cornell.edu}

\address[cornell]{Departments of Physics and Astronomy, Space
Sciences Building, Cornell University,
Ithaca, NY 14853, USA}

\address[caltech]{TAPIR, Walter Burke Institute for Theoretical Physics,
MC 350-17, California Institute of Technology,
Pasadena, CA 91125, USA}

\begin{abstract}
The DG algorithm is a powerful method for solving pdes, especially for
evolution equations
in conservation form. Since the algorithm involves integration
over volume elements, it is not immediately obvious that it will
generalize easily to arbitrary time-dependent curved spacetimes.
We show how to formulate the algorithm
in such spacetimes for
applications in relativistic astrophysics. We also show how to
formulate the algorithm for equations in non-conservative form,
such as Einstein's field equations themselves. We find two
computationally distinct formulations in both cases, one of which has seldom
been used  before for flat space in curvilinear coordinates but which
may be more efficient.
We also give a new derivation of the ALE algorithm
(Arbitrary Lagrangian-Eulerian) using 4-vector methods that
is much simpler than the usual derivation
and explains why the method preserves
the conservation form of the equations.
The various formulations are explored with some simple numerical experiments
that also explore the effect of the metric identities on the results.
\end{abstract}


\end{frontmatter}

\section{Introduction}
\label{sec:intro}
In relativistic astrophysics,
simulations involving hydrodynamics or magnetohydrodynamics or similar physics
are most often carried out using
finite-volume methods. Two major challenges of such simulations are
\emph{accuracy} and \emph{computational efficiency}.
Many important problems cannot be solved to the
required accuracy using currently available hardware resources.
Accuracy can be improved only by increasing numerical
resolution.
If parts of the solution are smooth so that one might want to
take advantage of
high-order methods to improve the accuracy, current methods eventually
run into problems.
High-order finite-volume methods couple together
more cells and require more communication between cells. Ultimately,
when the number of cells and processors gets large enough, the
communication time begins to limit the computation and the code no
longer scales with the number of processors.
Moreover, astrophysical applications often involve multiphysics
(fluids, magnetic fields, neutrinos, electromagnetic radiation,
relativistic gravity). With current formulations,
each new type of physics often requires
its own computational treatment, making coupling of the physics
difficult.
As one looks ahead to the arrival of exascale computing, it
seems that we should look to the development of algorithms that
can take advantage of these very large machines properly for astrophysics. 

In the last decade, discontinuous
Galerkin (DG) methods have emerged as the leading contender to
achieve all the goals of a general purpose simulation code,
particularly for equations in conservation form:
high order accuracy in smooth regions, robustness for shocks
and other discontinuities, scalability to very large machines, accurate
handling of irregular boundaries, adaptivity, and so on.
Many applications of DG in terrestrial fluid dynamics have appeared.
However, applications in relativity and astrophysics have so far been mainly
exploratory\cite{field:09,zumbusch2009,field:10,radice:11,brown:12dg,mocz:14,zanotti:14,endeve:15,zanotti2015,schaal2015}
and confined to simple problems.

The goal of this paper is to formulate the DG method
for arbitrary 3-dimensional problems
involving general relativistic gravitation.
At first sight, this sounds tricky: the basic DG algorithm involves
integrating the pdes over space and using Gauss's Theorem to turn
integrals of divergences into surface integrals. In general relativity,
spacetime is curved and coordinates are arbitrary and not necessarily
simply related to physical measurements by an observer.
So should integrations be performed over
coordinate volume or proper volume? What are the corresponding normal
vectors that enter into the interface flux prescriptions? How should
Einstein's equations, which are not typically in conservation form,
be handled? Is the weak form or the strong form of the equations
better? How do the so-called metric identities affect the formulation?
We give answers to these questions.
In particular, we find that the final formulation is very close to that
already developed for Euclidean space in curvilinear coordinates.
Moreover, the covariant approach adopted in this paper gives new
insights into the standard curvilinear coordinate treatment.
Not only are derivations much simpler, but alternative formulations
that may be more efficient computationally are found.

Here we summarize the key results in this paper.
\begin{itemize}
\item
Despite the curvature of spacetime, the DG algorithm can easily be formulated
in general relativity. In fact, the formulation is analogous to
that for curvilinear coordinates in flat spacetime.
\item
In the general case,
there are two distinct strong formulations for conservation laws.
For the tensor-product basis functions used in this paper, the
corresponding weak formulations are both equivalent to one of
the strong formulations.
\item
Only one of the formulations has been widely used for flat space in
curvilinear coordinates. In numerical experiments, the other appears
to be somewhat more efficient and should be further investigated.
\item
Similarly, there are two inequivalent formulations for hyperbolic
equations in non-conservation form. These formulations are important
for solving Einstein's equations.
\item
Time-dependent mappings (Arbitrary Lagrangian-Eulerian (ALE)
methods and the dual-frame approach\cite{scheel2006}
that has proved useful for black hole simulations) are easily
implemented in the relativistic treatment.
\item
We give streamlined derivations of the so-called metric identities,
the geometric conservation law, and the ALE method for moving
grids. The derivation of the ALE method
is novel and uses general covariance to get the result in a few lines.
In addition,
the reason that the ALE method preserves the conservation form of the
equations is explained.
\item
Satisfying the metric identities discretely is often claimed to be
a necessary condition for ``free-stream preservation,'' or the
requirement that a uniform flow remain uniform for all time.
We show that in fact this statement is true for only one of the
computational formulations of the DG algorithm and not the other.
\item
We clarify how normal vectors should be normalized. The normal
vector that the boundary flux vector is projected along does not
need to be the unit normal---the normalization factor cancels
out of the algorithm.
\end{itemize}

A covariant treatment of DG in general relativity has previously been
given by Radice and Rezzolla\cite{radice:11}. This paper covers many
aspects that were not covered by them.

\section{DG for equations in conservation form}
\label{sec:cons}
\subsection{Form of the equations}
In a general time-dependent curved spacetime, a conservation law can
be written in terms of a 4-divergence:
\begin{equation}
\nabla_\mu F^\mu = 0,
\label{eq:cons}
\end{equation}
where $\nabla_\mu$ denotes the covariant derivative.
Here and throughout, repeated indices are summed over. Greek indices
$\mu,\nu,\ldots$ range
from 0 to 3, while Latin indices $a,b,\ldots$
will be purely spatial, ranging from 1 to 3.
We choose units with the speed of light $c=1$, so that $x^0=t$.
We will often denote $F^0$ by $u$, a quantity like density that is conserved.
The spatial flux vector $F^a$ is generally a function of $u$.
In practice, rather than a single conservation law like (\ref{eq:cons}),
one deals with a system of conservation laws. In this case,
$u$ is a vector of conserved quantities and $F^a$ is a vector of flux
vectors.
For example, $u$ and $F^a$ are vectors of length 5 for hydrodynamics.
In this paper, we
will typically not need to deal with the various separate equations in a
system of conservation laws. Accordingly, we will write $u$ and $F^a$
whether we are dealing with one equation or a system. All the derivations
go through independently on each equation in a system.

It will be convenient to generalize (\ref{eq:cons}) to allow a source term
$s$ on the right-hand side. Such a source term arises, for example,
when one considers conservation of energy and momentum in a general curved
(or curvilinear)
metric. The divergence of the energy-momentum tensor gives an extra term
that cannot be included as the pure divergence of a flux vector.
However, the extra term depends only on $u$ and not on its
derivatives.  This is the  key requirement that we place on the
source term in the subsequent treatment.

The metric in a general spacetime can be written in the standard $3+1$ form
\begin{equation}
ds^2=g_{\mu\nu}dx^\mu dx^\nu
=-\alpha^2 dt^2+\gamma_{ab}(dx^a+\beta^a dt)(dx^b +\beta^b dt),
\label{eq:3+1}
\end{equation}
where $\alpha$ is called the lapse function, $\beta^a$ the shift vector,
and $\gamma_{ab}$ is the spatial metric on $t=\text{constant}$ slices.
(For more on the $3+1$ decomposition, see, e.g., \cite{baumgarte2010} or
\cite{rezzolla2013}.)
In flat spacetime (no relativistic gravitational field), we can
set $\alpha=1$, $\beta=0$. In Cartesian coordinates, $\gamma_{ab}=
\delta_{ab}$, the Euclidean 3-metric.

Now using a standard identity for the covariant
divergence (see, e.g., Problem 7.7(g) in \cite{lightman1975}),
the conservation equation (\ref{eq:cons}) with
source term can be written
\begin{equation}
\frac{1}{\sqrt{-g}}\partial_t(\sqrt{-g}u)
+
\frac{1}{\sqrt{-g}}\partial_a(\sqrt{-g}F^a) = s,
\label{eq:cons2}
\end{equation}
where $g$ is the determinant of the 4-metric $g_{\mu\nu}$. It is easy to
show from (\ref{eq:3+1})
that $\sqrt{-g}=\alpha\sqrt{\gamma}$, where $\gamma$ is
the determinant of $\gamma_{ab}$. So Eq.\ (\ref{eq:cons2}) becomes
\begin{equation}
\frac{1}{\sqrt{\gamma}}\partial_t(\sqrt{\gamma}u)
+
\frac{1}{\sqrt{\gamma}}\partial_a(\sqrt{\gamma}F^a) = s.
\label{eq:cons3}
\end{equation}
Here we have absorbed the factor of $\alpha$ in the definitions of $u$,
$F^a$, and $s$, which is standard practice in computational relativity.
Multiplying Eq.\ (\ref{eq:cons3}) through by $\sqrt{\gamma}$ gives
\begin{equation}
\partial_t(\sqrt{\gamma}u)
+
\partial_a(\sqrt{\gamma}F^a) = \sqrt{\gamma}s,
\label{eq:cons4}
\end{equation}
which suggests that $\sqrt{\gamma}u$, $\sqrt{\gamma}F^a$, and $\sqrt{\gamma}s$
might be convenient variables to use in a code.

Note that Eq.\ (\ref{eq:cons3}) or (\ref{eq:cons4}) is formally
the same as a conservation
law in flat space written in curvilinear coordinates. In that case the metric
is simply the Euclidean metric transformed to the curvilinear coordinates,
whereas here we consider the more general case where the metric might be
the solution of Einstein's equations of general relativity in an arbitrary
coordinate system.

\subsection{The DG algorithm in curved spacetime}
\label{sec:dgcurved}
As we will see, the DG algorithm is remarkably similar in flat and curved
spacetimes. However, at key points there are important details that
need to be correctly implemented.

The algorithm starts by dividing the spatial domain into subdomains, often
called cells or elements. Each element is a mapping of a reference element
(triangle or square in 2-d, tetrahedron or cube in 3-d).
The mapped elements can have straight or curved sides.
(Here ``straight'' means having two constant spatial coordinates.)
In each element, the quantities
$u$, $F^a$, $s$, etc.\ (or equivalently $\sqrt{\gamma}u$, $\sqrt{\gamma}F^a$,
$\sqrt{\gamma}s$)
are each expanded in term of a set of basis functions, typically polynomials:
\begin{equation}
u(\xbf)=\sum_i u_i \phi_i(\xbf).
\label{eq:2}
\end{equation}
Analogous to spectral collocation methods,
we will work with a so-called nodal expansion, where 
Eq.\ (\ref{eq:2}) is an interpolation:
\begin{equation}
u(\xbf_i) = u_i,\qquad i=0,\ldots,N
\label{eq:2a}
\end{equation}
for some choice of nodes (grid points) $\xbf_i$.
This implies that $\phi_i(\xbf_j)=\delta_{ij}$. Such basis functions
are called cardinal functions. For polynomial basis functions, 
they will be simply Lagrange interpolating polynomials.

Get the DG equations by integrating Eq.~(\ref{eq:cons3}) in each
element with test functions that are the same as the basis functions:
\begin{equation}
\int\left[
\partial_t\left(\sqrt{\gamma} u\right) + \partial_a\left(
\sqrt{\gamma}F^a\right)-\sqrt{\gamma}s \right]
\phi_i(\xbf)\,d^3x=0.
\label{eq:3}
\end{equation}
Here we have integrated with respect to \emph{proper} volume
$\sqrt{\gamma} d^3x$, where
$d^3x$ is the coordinate volume element $dx\,dy\,dz$.
Had we started with the form (\ref{eq:cons4}), then we
would have integrated with respect to \emph{coordinate} volume instead.
The choice is designed to allow Gauss's Theorem to be invoked in
a simple way, as we now show.

Transform the spatial derivative term by integrating by parts:
\begin{align}
\int \partial_a\left(\sqrt{\gamma}F^a\right)\phi_i(\xbf)\,d^3x & =
\int \partial_a\left(\sqrt{\gamma}F^a\phi_i(\xbf)\right)\,d^3x
- \int \sqrt{\gamma}F^a\partial_a \phi_i(\xbf)\,d^3x\nonumber\\
&=\oint F^a n_a\phi_i\,d^2\Sigma
- \int \sqrt{\gamma}F^a\partial_a \phi_i(\xbf)\,d^3x.
\label{eq:3a}
\end{align}
Here $d^2\Sigma$ is the proper surface element of the cell and $n_a$ is
the unit outward normal (see \ref{app:gauss} for more on Gauss's Theorem and
normal vectors.)

With a formulation like (\ref{eq:3a}) in each element, there is no
connection between the elements.
The heart of the DG method is to replace $F^a$ in the surface term by
the numerical flux $F^{a*}$, a function of the fluxes on \emph{both} sides
of the interface:
\begin{equation}
\int \partial_a\left(\sqrt{\gamma}F^a\right)\phi_i(\xbf)\,d^3x \to
\oint F^{a*} n_a\phi_i\,d^2\Sigma
- \int \sqrt{\gamma}F^a\partial_a \phi_i(\xbf)\,d^3x.
\label{eq:4}
\end{equation}
This gives the weak form of the equations (no derivatives of $F^a$).
To get the strong form, undo the integration by parts on
the right-hand side of Eq.\ (\ref{eq:4}):
\begin{equation}
\int \partial_a\left(\sqrt{\gamma}F^a\right)\phi_i(\xbf)\,d^3x
\to
\oint (F^{a*}-F^a) n_a\phi_i\,d^2\Sigma
+\int \partial_a\left(\sqrt{\gamma}F^a\right)\phi_i(\xbf)\,d^3x.
\label{eq:5}
\end{equation}
So one question we will have to answer is: Should we use the
weak form (\ref{eq:4}) or the strong form (\ref{eq:5})?
We will return to this question in \S\ref{subsec:weak}.

\subsubsection{The mass matrix}

Evaluation of Eq.\ (\ref{eq:3}) now follows standard lines as in flat
spacetime.
The first term gives the mass matrix:
\begin{align}
\int\partial_t \left(\sqrt{\gamma} u\right)\phi_i(\xbf)\,d^3x
&=\frac{d}{dt}\sum_j \left(\sqrt{\gamma} u\right)_j\int \phi_j(\xbf)\phi_i(\xbf)\,d^3x\nonumber\\
&=\Mbf\cdot \frac{d\left(\sqrt{\gammabf} \ubf\right)}{dt}\Big|_i,
\label{eq:6}
\end{align}
where the mass matrix $\Mbf$ has components
\begin{equation}
M_{ij}=\int \phi_i(\xbf)\phi_j(\xbf)\,d^3x
\label{eq:7}
\end{equation}
and the dot in Eq.~(\ref{eq:6}) denotes matrix-vector multiplication.

Note that a product like $\sqrt{\gamma} u$ is expanded using the product
of the function values at the grid points for each factor:
\begin{equation}
\sqrt{\gamma} u=\sum_j\sqrt{\gamma_j} u_j\phi_j(\xbf).
\end{equation}
This is not exact in general unless $\gamma$ is constant,
and will introduce aliasing that may have to be dealt with
by filtering. We will handle all products in this way.
The notation
$\sqrt{\gammabf} \ubf$
is used to denote the vector whose elements
are $\sqrt{\gamma_j} u_j$.

Similarly, the last term in Eq.~(\ref{eq:3}) gives
\begin{equation}
-\int \left(\sqrt{\gamma}s\right)\phi_i(\xbf)\,d^3x=-\Mbf\cdot
\left(\sqrt{\gammabf}\sbf\right)|_i .
\label{eq:8}
\end{equation}

\subsubsection{The volume derivative term}
For the strong form,
the derivative term in Eq.~(\ref{eq:3}) gives the boundary
and volume terms as in Eq.~(\ref{eq:5}).
Note that all spatial derivatives can be rewritten using
the derivative matrices defined by
\begin{equation}
D^{(a)}_{ij}=\partial_a \phi_j(\xbf_i).
\label{eq:8b}
\end{equation}
Explicitly, for any quantity $f(\xbf)$ we have
\begin{equation}
f(\xbf)=\sum_k f_k\phi_k(\xbf)\quad\Rightarrow\quad\partial_a f(\xbf)
=\sum_k f_k\partial_a\phi_k(\xbf)\quad\Rightarrow\quad
\partial_a f(\xbf)|_j=\sum_k D^{(a)}_{jk}f_k.
\label{eq:8c}
\end{equation}
Applying this to the derivative in Eq.~(\ref{eq:5}) gives
\begin{equation}
\partial_a\left(\sqrt{\gamma}F^a\right)=
\sum_j \left[\partial_a\left(\sqrt{\gamma}F^a\right)\right]_j\phi_j(\xbf)
=\sum_{jk} D^{(a)}_{jk}\left(\sqrt{\gamma}F^a\right)_k\phi_j(\xbf).
\label{eq:8a}
\end{equation}
Thus the volume term becomes
\begin{align}
\int \partial_a\left(\sqrt{\gamma}F^a\right)\phi_i(\xbf)\,d^3x
&=
\sum_{jk} D^{(a)}_{jk}\left(\sqrt{\gamma}F^a\right)_k
\int \phi_j(\xbf)\phi_i(\xbf)\,d^3x\nonumber\\
&=\sum_{jk}M_{ij} D^{(a)}_{jk}\left(\sqrt{\gamma}F^a\right)_k\nonumber\\
&=\Mbf\cdot \Dbf^{(a)}\cdot \left(\sqrt{\gammabf}\,\Fbf^a\right)\big|_.
\label{eq:9}
\end{align}

\subsubsection{The boundary flux}
For the boundary surface term, define
\begin{equation}
F\equiv (F^{a*}-F^a) n_a.
\label{eq:11}
\end{equation}
For smooth problems the numerical flux is generally evaluated by
upwinding.
For non-smooth solutions, the flux prescription typically enforces
the Rankine-Hugoniot relations in some way.
So let's assume we have some prescription for
$F$ on the boundary and that this has been computed as an expansion
in terms of the basis functions. Then
\begin{equation}
\oint F\phi_i\,d^2\Sigma=\sum_j F_j\oint \phi_j(\xbf)\phi_i(\xbf)\,\
d^2\Sigma.
\label{eq:10}
\end{equation}
Write this as
\begin{equation}
\oint F\phi_i\,d^2\Sigma=
\sum_{\rm surfaces}\Mbf^{(2)}\cdot\Fbf\big|_i,
\label{eq:10a}
\end{equation}
where the surface mass matrix on each piece of the boundary surface is
defined as
\begin{equation}
M^{(2)}_{ij}=\int\limits_{\rm surface}  \phi_i(\xbf)\phi_j(\xbf)\,d^2\Sigma.
\label{eq:10b}
\end{equation}

\subsection{The equations of motion}
Now put all the pieces together: Substitute Eqs.\ (\ref{eq:6}),
(\ref{eq:8}), (\ref{eq:9}) and (\ref{eq:10a}) in Eq.\ (\ref{eq:3}) and
multiply through by $\Mbf^{-1}$ to get
\begin{equation}
\frac{d\left(\sqrt{\gammabf} \ubf\right)}{dt}
+ \Dbf^{(a)}\cdot \left(\sqrt{\gammabf}\,\Fbf^a\right)
-\left(\sqrt{\gammabf}\sbf\right)
=-\Mbf^{-1}\cdot\sum_{\rm surfaces}\Mbf^{(2)}\cdot\Fbf.
\label{eq:10c}
\end{equation}
This is the form that is integrated with a suitable time stepper.
We see that the algorithm resembles the method of lines in finite-volume
methods.

\section{Evaluation of Integrals}
\label{sec:tensorproduct}
Evaluate the various integrals by mapping them to the reference element
and then doing a Gaussian quadrature.
Let the mapping be some time-independent function
\begin{equation}
\xbf=\xbf(\xbar)
\label{eq:12}
\end{equation}
with Jacobian matrix
\begin{equation}
\Jbf=\frac{\partial x^a}{\partial x^{\bar a}}
\label{eq:13}
\end{equation}
and Jacobian
\begin{equation}
J=\det\Jbf.
\label{eq:14}
\end{equation}
Here the barred coordinates are some standard coordinates covering the
reference element, which we will sometimes also denote as
\begin{equation}
(x^{\bar 1},x^{\bar 2},x^{\bar 3})=(a^1,a^2,a^3).
\label{eq:14a}
\end{equation}
Note the standard relativist's
convention of placing the bar on the index: $x^{\bar a}$
in (\ref{eq:13}) rather than
$\bar x^a$, which facilitates working with tensor indices.

We will take the reference element to be a cube
with extents $[-1,1]$ in each direction. The boundaries in
most typical astrophysics problems are sufficiently simple that
we don't need the flexibility of grids composed of triangles or
tetrahedra. Even for binary black hole simulations, where the centers
of each black hole are excised from the computational domain, we only
need a domain bounded by a large sphere at infinity with two interior
holes removed.
Such a domain can easily be covered with elements that are
mapped cubes.

Using cubes allows a big simplification in the algorithm:
We can take the basis functions to be a tensor
product of 1-d basis functions $\ell_i$ on the reference element:
\begin{equation}
\phi_i(\xbf)\to\phi_{ijk}(\xbf)=\ell_i\big(x^{\bar 1}\big)
\ell_j\big(x^{\bar 2}\big)\ell_k\big(x^{\bar 3}\big).
\label{eq:400}
\end{equation}
Since the Gaussian quadrature along each dimension
is being carried out with a weight function of unity,
it should be a Gauss-Legendre quadrature, with nodes $\bar x_i$
corresponding to roots of an appropriate Legendre polynomial.
A further simplification follows if we
assume that these
quadrature points are chosen to be the same as the interpolation
points.
The combination of tensor product basis functions
and quadrature points the same as interpolation points
gives a very simple form of the equations
that resembles multipenalty collocation.
It also implies that the basis functions are the Lagrange
interpolating polynomials
corresponding to Legendre polynomials:
\begin{equation}
\ell_i(\bar x) = \prod_{\substack{j=0\\j\neq i}}^N
\frac{\bar x-\bar x_j}{\bar x_i-\bar x_j}.
\end{equation}

Note that we can have different numbers of points along each dimension
in (\ref{eq:400}), which will lead to different values for the
quadrature points $\bar x_i$ and weights $w_i$ along each dimension.
We will not clutter the notation to make this distinction.
Also, we now have to give up the nice matrix notation we used above, since
each matrix index becomes a triple like $(ijk)$.

\subsection{The mass matrix}
Equation (\ref{eq:7}) for the mass matrix becomes
\begin{align}
M_{ij}\to M_{(ijk)(lmn)}&=\int \phi_{ijk}(\xbf)\phi_{lmn}(\xbf)
\,d^3x\nonumber\\
&=\int \ell_i\big(x^{\bar 1}\big)\ell_j\big(x^{\bar 2}\big)\ell_k\big(x^{\bar 3}\big)
  \ell_l\big(x^{\bar 1}\big)\ell_m\big(x^{\bar 2}\big)\ell_n\big(x^{\bar 3}\big)
  J(\xbar)\,d^3\bar x\nonumber\\
&=\sum_{pqr}
  \ell_i\big(x^{\bar 1}_p\big)
  \ell_l\big(x^{\bar 1}_p\big)w_p
  \ell_j\big(x^{\bar 2}_q\big)
  \ell_m\big(x^{\bar 2}_q\big)w_q
  \ell_k\big(x^{\bar 3}_r\big)
  \ell_n\big(x^{\bar 3}_r\big)w_r
  J\big(x^{\bar 1}_p,x^{\bar 2}_q,x^{\bar 3}_r\big)
  \nonumber\\
&=\sum_{pqr}\delta_{ip}\delta_{lp}w_p
  \delta_{jq}\delta_{mq}w_q
  \delta_{kr}\delta_{nr}w_r
  J_{pqr}\nonumber\\
&=\delta_{il}\delta_{jm}\delta_{kn}w_i w_j w_k J_{ijk}.
\label{eq:401}
\end{align}
Here we have used the equality of quadrature and interpolation points
to set quantities like $\ell_i\big(x^{\bar 1}_p\big)$ equal to $\delta_{ip}$.
The resulting mass matrix is diagonal in each dimension.

The ``matrix'' multiplication of the expression (\ref{eq:401}) by
a term like $d(\sqrt{\gamma}u)/dt|_{lmn}$ gives
\begin{equation}
w_i w_j w_k J_{ijk}\frac{d(\sqrt{\gamma}u)_{ijk}}{dt},
\label{eq:402}
\end{equation}
and a similar term for Eq. (\ref{eq:8}).

\subsection{The volume derivative term}
Corresponding to the manipulations in Eq.\ (\ref{eq:8c}) we have
\begin{align}
f(\xbf)=\sum_{ijk}f_{ijk}\phi_{ijk}(\xbf)
\quad&\Rightarrow\quad
\partial_a f(\xbf)= \sum_{ijk}f_{ijk}\partial_a \phi_{ijk}(\xbf)\nonumber\\
&\Rightarrow\quad
\partial_a f(\xbf)|_{lmn}=\sum_{ijk}f_{ijk}\partial_a
  \phi_{ijk}(\xbf_{lmn}).
\label{eq:403}
\end{align}
Mapping to the reference element gives
\begin{align}
\partial_a \phi_{ijk}(\xbf_{lmn}) &=
\frac{\partial x^{\bar a}}{\partial x^a}\partial_{\bar a}
\phi_{ijk}(\xbar)\big|_{lmn}\nonumber\\
&=
\frac{\partial x^{\bar a}}{\partial x^a}\partial_{\bar a}
\ell_i\big(x^{\bar 1}\big)\ell_j\big(x^{\bar 2}\big)\ell_k\big(x^{\bar 3}\big)
\big|_{lmn}\nonumber\\
&=
\frac{\partial x^{\bar 1}}{\partial x^a}D_{li}^{\bar 1}\delta_{jm}\delta_{kn}
+
\frac{\partial x^{\bar 2}}{\partial x^a}D_{mj}^{\bar 2}\delta_{il}\delta_{kn}
+
\frac{\partial x^{\bar 3}}{\partial x^a}D_{nk}^{\bar 3}\delta_{il}\delta_{jm},
\label{eq:404}
\end{align}
where
\begin{equation}
D_{li}^{\bar 1}=\partial_{\bar 1} \ell_i\big(x^{\bar 1}\big)\big|_l
\label{eq:405}
\end{equation}
is the derivative matrix for $x^{\bar 1}$, and similarly for the 2- and
3-coordinates.
Substituting Eq.\ (\ref{eq:404}) in Eq.\ (\ref{eq:403}) gives
\begin{equation}
\partial_a f(\xbf)|_{lmn}=
\frac{\partial x^{\bar 1}}{\partial x^a}\Big|_{lmn}\sum_i D_{li}^{\bar 1}f_{imn}
+
\frac{\partial x^{\bar 2}}{\partial x^a}\Big|_{lmn}\sum_j D_{mj}^{\bar 2}f_{ljn}
+
\frac{\partial x^{\bar 3}}{\partial x^a}\Big|_{lmn}\sum_k D_{nk}^{\bar 3}f_{lmk}
.
\label{eq:406}
\end{equation}
So analogously to Eq.\ (\ref{eq:8a}) we get
\begin{equation}
\partial_a\left(\sqrt{\gamma}F^a\right)=
\sum_{lmn}\left[\partial_a\left(\sqrt{\gamma}F^a\right)\right]_{lmn}
\phi_{lmn}(\xbf),
\label{eq:407}
\end{equation}
where $[\partial_a\left(\sqrt{\gamma}F^a\right)]_{lmn}$
is given by the analog of Eq.\ (\ref{eq:406}). Thus the
volume term (\ref{eq:9}) gives
\begin{align}
\int \partial_a\left(\sqrt{\gamma}F^a\right)\phi_i(\xbf)\,d^3x
&=\sum_{lmn}\left[\partial_a\left(\sqrt{\gamma}F^a\right)\right]_{lmn}
  \int\phi_{lmn}(\xbf)\phi_{ijk}(\xbf)\,d^3x\nonumber\\
&=\sum_{lmn}\left[\partial_a\left(\sqrt{\gamma}F^a\right)\right]_{lmn}
  M_{(ijk)(lmn)}\nonumber\\
&=\left[\partial_a\left(\sqrt{\gamma}F^a\right)\right]_{ijk}w_i w_j w_k J_{ijk},
\label{eq:408}
\end{align}
where we have used Eq.\ (\ref{eq:401}) in the last line.

\subsection{The boundary flux}
Let's consider the flux through the surface corresponding to $a^3=
x^{\bar 3}=1$
on the reference element.
Evaluating Eq.\ (\ref{eq:10}) by transforming
to the reference element gives
\begin{equation}
\int F \phi_{ijk}(\xbf)\,d^2\Sigma =
\sum_{lmn}F_{lmn}\int\limits_{a^3=1}
 \ell_l\big(x^{\bar 1}\big)\ell_m\big(x^{\bar 2}\big)\ell_n\big(x^{\bar 3}\big)
  \ell_i\big(x^{\bar 1}\big)\ell_j\big(x^{\bar 2}\big)\ell_k\big(x^{\bar 3}\big)
  \sqrt{{}^{(2)}\gamma\big(x^{\bar 1},x^{\bar 2}\big)}\,
  dx^{\bar 1}\,dx^{\bar 2}.
\label{eq:409}
\end{equation}
Here ${}^{(2)}\gamma$ is the determinant of the
2-dimensional metric induced on the surface by $\gamma_{ij}$.

Assume the quadrature uses Gauss-Lobatto points. Then $\ell_k\big(x^{\bar 3}
\big)=\delta_{kN}$, where $N$ is the last grid point. Equation (\ref{eq:409})
becomes
\begin{align}
\int F \phi_{ijk}(\xbf)\,d^2\Sigma &=
\sum_{lmn,pq}F_{lmn}\sqrt{{}^{(2)}\gamma_{pq}}\,
\delta_{lp}\delta_{mq}\delta_{nN}\delta_{ip}
\delta_{jq}\delta_{kN}w_p w_q\nonumber\\
&=F_{ijN}\sqrt{{}^{(2)}\gamma_{ij}}\, w_i w_j \delta_{kN}.
\label{eq:410}
\end{align}

Integrating over the surface $a^3=-1$ gives a similar term with a
minus sign and $N$ replaced
by the index 0, corresponding to the first grid point in the interval.
The contributions from the other 4 sides of the cube are similar.

\subsection{The equations of motion --- strong form}
Putting together Eqs.\ (\ref{eq:402}), (\ref{eq:408}), (\ref{eq:406}) and
(\ref{eq:410}), and dividing through
by $w_i w_j w_k J_{ijk}$, we get
\begin{align}
\frac{d(\sqrt{\gamma}u)_{ijk}}{dt}  &+
\Big[
\frac{\partial x^{\bar 1}}{\partial x^a}\Big|_{ijk}\sum_l D_{il}^{\bar 1}
\left(\sqrt{\gamma}F^a\right)_{ljk}
+
\frac{\partial x^{\bar 2}}{\partial x^a}\Big|_{ijk}\sum_m D_{jm}^{\bar 2}
\left(\sqrt{\gamma}F^a\right)_{imk}\nonumber\\
&\quad+
\frac{\partial x^{\bar 3}}{\partial x^a}\Big|_{ijk}\sum_n D_{kn}^{\bar 3}
\left(\sqrt{\gamma}F^a\right)_{ijn}
\Big]-(\sqrt{\gamma}s)_{ijk}\nonumber\\
&=-
\frac{1}{w_N}F_{ijN}\frac{\sqrt{{}^{(2)}\gamma_{ij}}}{J_{ijN}}
 \delta_{kN}
+
\frac{1}{w_0}F_{ij0}\frac{\sqrt{{}^{(2)}\gamma_{ij}}}{J_{ij0}}
 \delta_{k0}
-
\frac{1}{w_N}F_{Njk}\frac{\sqrt{{}^{(2)}\gamma_{jk}}}{J_{Njk}}
 \delta_{iN}\nonumber\\
&\quad+
\frac{1}{w_0}F_{0jk}\frac{\sqrt{{}^{(2)}\gamma_{jk}}}{J_{0jk}}
 \delta_{i0}
-
\frac{1}{w_N}F_{iNk}\frac{\sqrt{{}^{(2)}\gamma_{ik}}}{J_{iNk}}
 \delta_{jN}
+
\frac{1}{w_0}F_{i0k}\frac{\sqrt{{}^{(2)}\gamma_{ik}}}{J_{i0k}}
 \delta_{j0}.
\label{eq:411}
\end{align}
The relatively simple form for the boundary flux terms in Eq.\ (\ref{eq:411})
occurs because
of the 1-d Gauss-Lobatto quadratures using the interpolation
points.\footnote{If we had used Gauss interior
points instead of Gauss-Lobatto points,
then an interpolation would be required to evaluate quantities
on the boundary. Several works have compared the relative advantages
of these choices of quadrature points (see \cite{kopriva2010} and references
therein). However, these comparisons have typically been done with
relatively simple fluxes and no source terms. Since we have in mind
relativistic applications where complicated source terms
and equation of state calculations could dominate the
computational time, we will consider only the simpler Lobatto choice here.}

Note that the flux terms can be rewritten using Eq.\ (\ref{eq:211}).
For example, for the surface $x^{\bar 3} =$ constant we get
\begin{align}
\big(F^{a*}-F^a\big) n_a\frac{\sqrt{{}^{(2)}\gamma}}{J}&=
\big(F^{a*}-F^a\big)\left(
\frac{\partial x^{\bar 3}}{\partial x^a}J
 \frac{\sqrt{\gamma}}{\sqrt{{}^{(2)}\gamma}} \right)
\frac{\sqrt{{}^{(2)}\gamma}}{J}\nonumber\\
&=\sqrt{\gamma}\big(F^{a*}-F^a\big) \frac{\partial x^{\bar 3}}{\partial x^a},
\label{eq:412}
\end{align}
and similarly for the other surfaces. In other words, the ``normal vector''
can be taken to be simply $n_a=\partial x^{\bar 3}/\partial x^a$ since
the factor $\sqrt{\gamma}$ is usually incorporated in the definition of
$F^a$.
The simplified version of the right-hand side of Eq.\ (\ref{eq:411})
is
\begin{align}
{}-&
\frac{1}{w_N}\big[\sqrt{\gamma}\big(F^{a*}-F^a\big)\big]_{ijN}
 \frac{\partial x^{\bar 3}}{\partial x^a}\Big|_{ijN}
 \delta_{kN}
+
\frac{1}{w_0}\big[\sqrt{\gamma}\big(F^{a*}-F^a\big)\big]_{ij0}
 \frac{\partial x^{\bar 3}}{\partial x^a}\Big|_{ij0}
 \delta_{k0}
 \nonumber\\
{}-&
\frac{1}{w_N}\big[\sqrt{\gamma}\big(F^{a*}-F^a\big)\big]_{Njk}
 \frac{\partial x^{\bar 1}}{\partial x^a}\Big|_{Njk}
 \delta_{iN}
+
\frac{1}{w_0}\big[\sqrt{\gamma}\big(F^{a*}-F^a\big)\big]_{0jk}
 \frac{\partial x^{\bar 1}}{\partial x^a}\Big|_{0jk}
 \delta_{i0}
 \nonumber\\
{}-&
\frac{1}{w_N}\big[\sqrt{\gamma}\big(F^{a*}-F^a\big)\big]_{iNk}
 \frac{\partial x^{\bar 2}}{\partial x^a}\Big|_{iNk}
 \delta_{jN}
+
\frac{1}{w_0}\big[\sqrt{\gamma}\big(F^{a*}-F^a\big)\big]_{i0k}
 \frac{\partial x^{\bar 2}}{\partial x^a}\Big|_{i0k}
 \delta_{j0}.
\label{eq:bdy}
\end{align}
Note the symmetry between the Jacobian matrix elements that appear
on the left-hand side of Eq.\ (\ref{eq:411}) and in the boundary
flux terms Eq.\ (\ref{eq:bdy}), reflecting the essence of Gauss's
Theorem.

\subsection{The equations of motion --- weak form}

The weak form of the equations follows from using Eq.~(\ref{eq:4}) instead
of Eq.~(\ref{eq:5}). The boundary flux now contains only $F^{a*}$ 
instead of $(F^{a*}-F^a)$, while the volume term is now
\begin{align}
\text{Vol.\ term} &=
- \int \sqrt{\gamma}F^a\partial_a \phi_i(\xbf)\,d^3x\nonumber\\
&=
-\sum_j\left(\sqrt{\gamma}F^a\right)_j
\int \phi_j(\xbf)\partial_a \phi_i(\xbf)\,d^3x.
\label{eq:weak1}
\end{align}
The derivative of the basis function in Eq.~(\ref{eq:weak1}) is
\begin{equation}
\partial_a\phi_i(\xbf)=\sum_k\partial_a\phi_i(\xbf_k)\phi_k(\xbf)
=\sum_k D_{ki}^{(a)}\phi_k(\xbf)
\end{equation}
and so
\begin{align}
\text{Vol.\ term} &=
-\sum_{jk}\sqrt{\gamma_j}F^a_j D_{ki}^{(a)}\int\phi_j(\xbf)\phi_k(\xbf)\,d^3x
\nonumber\\
&=-\sum_{jk}\sqrt{\gamma_j}F^a_j M_{jk} D_{ki}^{(a)}\nonumber\\
&=-\left.\Dbf^{(a)\,T}\cdot\Mbf\cdot\left(\sqrt{\gammabf}\,\Fbf^a\right)\right|_i.
\end{align}

Now evaluate the volume term explicitly using Gaussian quadrature.
Equation (\ref{eq:weak1}) gives
\begin{align}
\text{Vol.\ term}
&=-\sum_{lmn}\left(\sqrt{\gamma}F^a\right)_{lmn}
  \int\phi_{lmn}(\xbf)\partial_a\phi_{ijk}(\xbf)\,d^3x\nonumber\\
&=-\sum_{lmn}\left(\sqrt{\gamma}F^a\right)_{lmn}
  \int\phi_{lmn}(\xbar)\frac{\partial x^{\bar a}}{\partial x^a}
  \partial_{\bar a}\phi_{ijk}(\xbar)J\,d^3\xbar\nonumber\\
&=-\sum_{lmn,pqr}\left(\sqrt{\gamma}F^a\right)_{lmn}
   \ell_l\big(x^{\bar 1}_p\big)w_p
   \ell_m\big(x^{\bar 2}_q\big)w_q
   \ell_n\big(x^{\bar 3}_r\big)w_r
   J_{pqr}\left.\frac{\partial x^{\bar a}}{\partial x^a}\right|_{pqr}
   \partial_{\bar a}
   \ell_i\big(x^{\bar 1}_p\big)
   \ell_j\big(x^{\bar 2}_q\big)
   \ell_k\big(x^{\bar 3}_r\big).
\label{eq:weak3}
\end{align}
The term corresponding to $x^{\bar a}=x^{\bar 1}$ is
\begin{align}
-&\sum_{lmn,pqr}\left(\sqrt{\gamma}F^a\right)_{lmn}
   \ell_l\big(x^{\bar 1}_p\big)w_p
   \ell_m\big(x^{\bar 2}_q\big)w_q
   \ell_n\big(x^{\bar 3}_r\big)w_r
   J_{pqr}\left.\frac{\partial x^{\bar 1}}{\partial x^a}\right|_{pqr}
   \partial_{\bar 1}
   \ell_i\big(x^{\bar 1}_p\big)
   \ell_j\big(x^{\bar 2}_q\big)
   \ell_k\big(x^{\bar 3}_r\big)
   \nonumber\\
{}=-&
\sum_{lmn,pqr}\left(\sqrt{\gamma}F^a\right)_{lmn}
\delta_{lp}
\delta_{mq}
\delta_{nr}
J_{pqr}\left.\frac{\partial x^{\bar 1}}{\partial x^a}\right|_{pqr}
D^{\bar 1}_{pi}
\delta_{jq}
\delta_{kr}
w_p w_q w_r
\nonumber\\
{}=-&
\sum_l \left(\sqrt{\gamma}F^a\right)_{ljk}
J_{ljk}\left.\frac{\partial x^{\bar 1}}{\partial x^a}\right|_{ljk}
D^{\bar 1}_{li}
w_l w_j w_k.
\label{eq:weak5}
\end{align}
We get similar terms for $x^{\bar 2}$ and $x^{\bar 3}$. Dividing through
by $w_i w_j w_k J_{ijk}$ gives a volume term that replaces the
term in square brackets in Eq.~(\ref{eq:411}):
\begin{align}
-&\frac{1}{J_{ijk}}\Big[
\sum_l \tilde D_{il}^{\bar 1}
\frac{\partial x^{\bar 1}}{\partial x^a}\Big|_{ljk}J_{ljk}
\left(\sqrt{\gamma}F^a\right)_{ljk}
+
\sum_m \tilde D_{jm}^{\bar 2}
\frac{\partial x^{\bar 2}}{\partial x^a}\Big|_{imk}J_{imk}
\left(\sqrt{\gamma}F^a\right)_{imk}
\nonumber\\
&\quad+
\sum_n \tilde D_{kn}^{\bar 3}
\frac{\partial x^{\bar 3}}{\partial x^a}\Big|_{ijn}J_{ijn}
\left(\sqrt{\gamma}F^a\right)_{ijn}
\Big],
\label{eq:weak4}
\end{align}
where $\tilde D$ is the differentiation matrix for the weak form:
\begin{equation}
\tilde D_{il} = \frac{w_l}{w_i}D_{li}.
\label{eq:weak_deriv}
\end{equation}

\section{Alternative Formulation: Transform Then Integrate}
\label{sec:transform}
In the above formulation of DG, we integrated the flux-conservative
equations against the test functions $\phi_i(\xbf)$ and then evaluated
the integrals by transforming to the reference grid $\xbar$. Instead
of this ``integrate then transform'' approach, we can transform first,
then integrate. Analytically, this is completely equivalent, but this is not
necessarily so in the discrete case.

Transforming Eq.\ (\ref{eq:cons4}) gives
\begin{equation}
\partial_t\left(\sqrt{\bar\gamma} \bar u\right) +
\partial_{\bar a}\left(
\sqrt{\bar\gamma}F^{\bar a}\right)=\sqrt{\bar\gamma}\bar s.
\label{eq:500}
\end{equation}
This form follows immediately by noting that Eq.\ (\ref{eq:cons4})
comes directly from the coordinate-independent expression (\ref{eq:cons}).
Now since the transformation Eq.\ (\ref{eq:12}) is independent of $t$,
we can consider just the spatial 3-vector properties of the transformation
and leave the time components of 4-vectors unchanged: $\bar u = u$,
$\bar s=s$.\footnote{
For the momentum equation, where $u\sim T^0{}_b$ is the
$b$-component of the momentum density,
we leave the $b$ index
untransformed, and $u$ remains the momentum component in the original frame.}
The determinant of the metric transforms according to Eq.\ (\ref{eq:209}),
and so Eq.\ (\ref{eq:500}) becomes
\begin{equation}
\partial_t\left(J\sqrt{\gamma} u\right) +
\partial_{\bar a}\left(
J\sqrt{\gamma}F^{\bar a}\right)=J\sqrt{\gamma}s,
\label{eq:501}
\end{equation}
where
\begin{equation}
F^{\bar a} = \frac{\partial  x^{\bar a}}{\partial x^a} F^a.
\label{eq:502}
\end{equation}
Integrating Eq.\ (\ref{eq:501}) against the basis functions in
the reference frame gives
\begin{equation}
\int\left[
\partial_t\left(J\sqrt{\gamma} u\right) + \partial_{\bar a}
\left(J\sqrt{\gamma}F^{\bar a}\right)-J\sqrt{\gamma}s \right]
\phi_i(\xbar)\,d^3\xbar=0.
\label{eq:503}
\end{equation}
This equation is equivalent to Eq.\ (\ref{eq:3}) provided
\begin{equation}
\partial_{\bar a}\left(J \frac{\partial  x^{\bar a}}{\partial x^a}\right) = 0,
\label{eq:504}
\end{equation}
as we see from Eq.\ (\ref{eq:502}).
Equation (\ref{eq:504}) is in fact a set of identities that
follows from symmetry properties of
the Jacobian matrix and is called the \emph{metric identities}.
For completeness, we give a derivation of Eq.\ (\ref{eq:504})
in \ref{app:identities}.
From that derivation, it is clear that the metric $\gamma_{ab}$ really has
nothing to do with the identities. Rather, in the
special case that the original metric is flat space
in Euclidean form, so that $\gamma_{ab}=\delta_{ab}$, then Eq.\ (\ref{eq:208})
shows that the transformed metric $\gamma_{\bar a\bar b}$ is given
by elements of the Jacobian matrix,
which is the origin of the name.

Now consider evaluating Eq.\ (\ref{eq:503}) by Gaussian quadrature. The first
term gives Eq.\ (\ref{eq:402}) again, with a similar term for the third
term. For the second term, we carry out the standard manipulations
that led to Eqs.~(\ref{eq:4}) and (\ref{eq:5}) to get
\begin{equation}
\everymath={\displaystyle}
\text{Vol.\ term} = \begin{cases}
\int\partial_{\bar a}\left(J\sqrt{\gamma}F^{\bar a}\right)\phi_i(\xbar)
  \,d^3\xbar, & \text{strong form}\\[5pt]
-\int J\sqrt{\gamma}F^{\bar a} \partial_{\bar a}\phi_i(\xbar)
  \,d^3\xbar, & \text{weak form}.
\end{cases}
\label{eq:505}
\end{equation}
The boundary term is
\begin{align}
\int\partial_{\bar a}\left(J\sqrt{\gamma}F^{\bar a}\phi_i(\xbar)\right)
  \,d^3\xbar
&=\oint \sqrt{\gamma} J\bar F \phi_i(\xbar)/\sqrt{\bar\gamma}\,d^2\Sigma
\nonumber\\
&=\oint \bar F \phi_i(\xbar)\,d^2\Sigma,
\label{eq:506}
\end{align}
which is the same as Eq.\ (\ref{eq:409}) since $\bar F = F$ is a scalar.
(Note that $d^2\Sigma$ is also an invariant.)

\subsection{Strong form}
Evaluating the strong form of the volume term (\ref{eq:505}) gives
\begin{align}
&\sum_{lmn}\big(J\sqrt{\gamma}F^{\bar a}\big)_{lmn}
  \int \phi_{ijk}(\xbar)
\partial_{\bar a}\phi_{lmn}(\xbar)\,d^3\bar x\nonumber\\
{}=&
\sum_{lmn,pqr}\big(J\sqrt{\gamma}F^{\bar a}\big)_{lmn}
   \ell_i\big(x^{\bar 1}_p\big)
   \ell_j\big(x^{\bar 2}_q\big)
   \ell_k\big(x^{\bar 3}_r\big)
   \partial_{\bar a}
   \ell_l\big(x^{\bar 1}_p\big)
   \ell_m\big(x^{\bar 2}_q\big)
   \ell_n\big(x^{\bar 3}_r\big)
   w_p w_q w_r.
\end{align}
The term corresponding to $x^{\bar a}=x^{\bar 1}$ is
\begin{align}
&\sum_{lmn,pqr}\big(J\sqrt{\gamma}F^{\bar 1}\big)_{lmn}
   \ell_i\big(x^{\bar 1}_p\big)
   \ell_j\big(x^{\bar 2}_q\big)
   \ell_k\big(x^{\bar 3}_r\big)
   \partial_{\bar 1}
   \ell_l\big(x^{\bar 1}_p\big)
   \ell_m\big(x^{\bar 2}_q\big)
   \ell_n\big(x^{\bar 3}_r\big)
   w_p w_q w_r\\
{}=&
\sum_{lmn,pqr}\big(J\sqrt{\gamma}F^{\bar 1}\big)_{lmn}
\delta_{ip}
\delta_{jq}
\delta_{kr}
D^{\bar 1}_{pl}
\delta_{mq}
\delta_{nr}
   w_p w_q w_r\\
{}=&
\sum_{l}\big(J\sqrt{\gamma}F^{\bar 1}\big)_{ljk}
D^{\bar 1}_{il}
w_i w_j w_k,
\end{align}
with similar terms for $x^{\bar 2}$ and $x^{\bar 3}$. Dividing through
by $w_i w_j w_k J_{ijk}$ gives the volume term that replaces the
term in square brackets in Eq.~(\ref{eq:411}):
\begin{equation}
\frac{1}{J_{ijk}}\Big[
\sum_{l} D^{\bar 1}_{il}
\big(J\sqrt{\gamma}F^{\bar 1}\big)_{ljk}
+
\sum_{m} D^{\bar 2}_{jm}
\big(J\sqrt{\gamma}F^{\bar 2}\big)_{imk}
+
\sum_{n} D^{\bar 3}_{kn}
\big(J\sqrt{\gamma}F^{\bar 3}\big)_{ijn}
\Big].
\label{eq:509}
\end{equation}
In comparing Eq.\ (\ref{eq:509}) with Eq.\ (\ref{eq:411}),
note that the term $\partial x^{\bar a}/\partial x^a$
is present in the definition of $F^{\bar a}$. The key result
is that this expression is different from the integrate-then-transform
result in that the $J\partial x^{\bar a}/\partial x^a$ term is now
being operated on by $D$ rather than being outside this operator.

\subsection{Weak form}
Similarly, evaluating the weak form of the volume term (\ref{eq:505}) gives
\begin{align}
-&\sum_{lmn}\big(J\sqrt{\gamma}F^{\bar a}\big)_{lmn}
  \int\phi_{lmn}(\xbar)\partial_{\bar a}\phi_{ijk}(\xbar)\,d^3\bar x\nonumber\\
{}= -&
\sum_{lmn,pqr}\big(J\sqrt{\gamma}F^{\bar a}\big)_{lmn}
   \ell_l\big(x^{\bar 1}_p\big)
   \ell_m\big(x^{\bar 2}_q\big)
   \ell_n\big(x^{\bar 3}_r\big)
   \partial_{\bar a}
   \ell_i\big(x^{\bar 1}_p\big)
   \ell_j\big(x^{\bar 2}_q\big)
   \ell_k\big(x^{\bar 3}_r\big)
   w_p w_q w_r.
\end{align}
The term corresponding to $x^{\bar a}=x^{\bar 1}$ is
\begin{align}
&-\sum_{lmn,pqr}\big(J\sqrt{\gamma}F^{\bar 1}\big)_{lmn}
   \ell_l\big(x^{\bar 1}_p\big)
   \ell_m\big(x^{\bar 2}_q\big)
   \ell_n\big(x^{\bar 3}_r\big)
   \partial_{\bar 1}
   \ell_i\big(x^{\bar 1}_p\big)
   \ell_j\big(x^{\bar 2}_q\big)
   \ell_k\big(x^{\bar 3}_r\big)
   w_p w_q w_r\\
{}=&-
\sum_{lmn,pqr}\big(J\sqrt{\gamma}F^{\bar 1}\big)_{lmn}
\delta_{lp}
\delta_{mq}
\delta_{nr}
D^{\bar 1}_{pi}
\delta_{jq}
\delta_{kr}
   w_p w_q w_r\\
{}=&-
\sum_{l}\big(J\sqrt{\gamma}F^{\bar 1}\big)_{ljk}
D^{\bar 1}_{li}
w_l w_j w_k.
\end{align}
This is identical to Eq.\ (\ref{eq:weak5}), so in the weak form 
integrate-then-transform gives the same result as transform-then-integrate.

\subsection{Equivalence of strong and weak forms for transform-then-integrate}
\label{subsec:weak}
So far, we have shown that the weak form of the equations is the same
whether we integrate first or transform first. However, the strong forms
are different from each other and appear different from the weak forms.
We now show that the strong form is algebraically identical to
the weak form for transform-then-integrate. This means that in a numerical
code it would differ only by roundoff and perhaps in efficiency.

The easiest way to see the equivalence is to use the following identity
for Gauss-Legendre-Lobatto differentiation matrices:
\begin{equation}
w_iD_{il}+w_lD_{li}=\delta_{Ni}\delta_{Nl}-\delta_{0i}\delta_{0l}.
\label{eq:derivmat}
\end{equation}
This identity is derived in \ref{app:derivmat}.
The identity allows us to convert weak differentiation matrices into
strong and vice versa, and it
shows that the weak form
(\ref{eq:weak4}) is identical
to the strong form (\ref{eq:509}) plus the boundary terms necessary to
change $F^{a*}$ into $(F^{a*}-F^a)$.

For the case of flat space, the equivalence between the strong and
weak forms of the equations
for transform-then-integrate has previously been shown 
by a different method in \cite{kopriva2010}.

To summarize:
\begin{itemize}
\item
The strong form of the equations gives rise to two computationally distinct
forms, corresponding to the integrate-then-transform and
transform-then-integrate approaches.
The first form is given by Eqs.\ (\ref{eq:411}) and (\ref{eq:bdy}). The second
form is given by Eq.\ (\ref{eq:509}).
\item
The weak form of the equations appears also to come in two forms. However,
both approaches lead to Eq.\ (\ref{eq:weak4}).
\item
The weak form is algebraically equivalent to the second strong form (\ref{eq:509}).
So in practice only the two strong forms need be considered.
\end{itemize}

The transform-then-integrate strong form (\ref{eq:509}), or the equivalent weak
form (\ref{eq:weak4}), is the formulation that has generally been used
previously for
hexadedral meshes, with $\sqrt{\gamma}=1$ because of flat space and
with the factor $J_{ijk}$ not divided out but kept with the variable $u$
in the time derivative
(see, e.g., the textbook \cite{kopriva2009}.) 
As we will see in the numerical experiments below, the alternative
form (\ref{eq:411}) may offer advantages in some cases.

Note that even if the metric identities are satisfied discretely,
the two strong forms of the algorithm are not equivalent. The reason is
that the derivative matrix does not in general satisfy the product rule for
derivatives. While the matrix gives the exact derivative for a polynomial
of degree $N$, a product is a polynomial of degree $2N$.

Note also that if we set $\sqrt{\gamma}F^a={}$constant in Eq.\ (\ref{eq:509}),
then the time derivative will vanish if the metric identities are
satisfied discretely. (Here we assume that the boundary conditions
do not spoil this statement.) This fact is the basis for the claim that
satisfying the metric identities discretely is
a necessary condition for ``free-stream preservation,'' or the
requirement that a uniform flow remain uniform for all time.
However, for the alternative form (\ref{eq:411}),
the time derivative vanishes when $\sqrt{\gamma}F^a={}$constant
without any requirement on the metric identities.

Finally, note that when the mapping from the physical frame to the reference
frame is linear, as in the common case of a Cartesian grid with elements
simply translated in $x$, $y$, and $z$, the elements of the Jacobian
matrix are constant and the two computational forms discussed here are
the same. Another special case where there is only one distinct
formulation is in one dimension. The reason is that the Jacobian determinant
and the Jacobian matrix are identical in this case.

\section{Equations in non-conservative form --- Einstein's equations}
The two most widely used formulations of Einstein's equations for
numerical work are the Generalized Harmonic equations in first-order
form \cite{lindblom2006}
and the BSSN formulation \cite{shibata1995,baumgarte1999}.
Neither of these is in conservation
form.
However, since these formulations
of Einstein's equations do not lead to shocks in the gravitational field
variables, it is not clear that there is any advantage in trying to find
flux-conservative forms. Accordingly, we need to develop a DG algorithm
for hyperbolic equations of the form
\begin{equation}
\partial_t u + A^a\partial_a u=s,
\label{eqq:1}
\end{equation}
where $A$ and $u$ are sufficiently smooth.\footnote{When $u$ can
be discontinuous the algorithm becomes much more complicated and is
beyond the scope of this paper.}
Again, everything in this section
goes through if (\ref{eqq:1}) is a system of equations, with
$A^a$ a set of matrices multiplying the vector $u$.

Get the DG equations by integrating Eq.~(\ref{eqq:1}) with the
basis functions over proper
volume\footnote{We could
integrate over coordinate
volume instead. Then one uses Gauss's Theorem in the form (\ref{eq:201}).
When using tensor product basis functions, the final expression
(\ref{eqq:411}) turns out to be unchanged.}:
\begin{equation}
\int\left(\partial_t u + A^a\partial_a u-s\right)
\phi_i(\xbf)\sqrt{\gamma}\,d^3x=0.
\label{eqq:3}
\end{equation}
Transform the spatial derivative term in the usual way:
\begin{align}
\int A^a\partial_a u\,\phi_i(\xbf)\sqrt{\gamma}\,d^3x
&\to
\oint (A^a u)^* n_a\phi_i\,d^2\Sigma
-\int u\partial_a (A^a \phi_i\sqrt{\gamma})\,d^3x\label{eqq:4}\\
&=\oint [(A^a u)^*-A^a u] n_a\phi_i\,d^2\Sigma
+\int A^a \partial_a u\,\phi_i\sqrt{\gamma}\,d^3x.
\label{eqq:5}
\end{align}
As expected, the quantity $A^a u$ plays the role of the flux.
Using the strong form (\ref{eqq:5}) and
carrying out the expansion in basis functions as before, we find
the equation analogous to Eq.\ (\ref{eq:10c}):
\begin{equation}
\frac{d\ubf}{dt}+\Abf^a\cdot \Dbf^{(a)}\cdot \ubf-\sbf=-
\Mbf^{-1}\cdot\sum_{\rm surfaces}\Mbf^{(2)}\cdot\Fbf.
\label{eqq:10c}
\end{equation}
Here $F=[(A^a u)^*-A^a u]n_a$ is the boundary flux for the strong form
of the equations.
\subsection{Non-conservative equations --- strong form}
Carrying out the integrals in (\ref{eqq:10c}) by Gauss-Legendre-Lobatto
quadrature as before, we get the equations of motion
\begin{align}
\frac{du_{ijk}}{dt}&+A^a_{ijk}
\Big[
\frac{\partial x^{\bar 1}}{\partial x^a}\Big|_{ijk}\sum_l D_{il}^{\bar 1}u_{ljk}
+
\frac{\partial x^{\bar 2}}{\partial x^a}\Big|_{ijk}\sum_m D_{jm}^{\bar 2}u_{imk}
+
\frac{\partial x^{\bar 3}}{\partial x^a}\Big|_{ijk}\sum_n D_{kn}^{\bar 3}u_{ijn}
\Big]-s_{ijk}\nonumber\\
=-&
\frac{1}{w_N}\big[\big(A^{a}u\big)^*-A^a u\big]_{ijN}
 \frac{\partial x^{\bar 3}}{\partial x^a}\Big|_{ijN}
 \delta_{kN}
+
\frac{1}{w_0}\big[\big(A^{a}u\big)^*-A^a u\big]_{ij0}
 \frac{\partial x^{\bar 3}}{\partial x^a}\Big|_{ij0}
 \delta_{k0}
 \nonumber\\
{}-&
\frac{1}{w_N}\big[\big(A^{a}u\big)^*-A^a u\big]_{Njk}
 \frac{\partial x^{\bar 1}}{\partial x^a}\Big|_{Njk}
 \delta_{iN}
+
\frac{1}{w_0}\big[\big(A^{a}u\big)^*-A^a u\big]_{0jk}
 \frac{\partial x^{\bar 1}}{\partial x^a}\Big|_{0jk}
 \delta_{i0}
 \nonumber\\
{}-&
\frac{1}{w_N}\big[\big(A^{a}u\big)^*-A^a u\big]_{iNk}
 \frac{\partial x^{\bar 2}}{\partial x^a}\Big|_{iNk}
 \delta_{jN}
+
\frac{1}{w_0}\big[\big(A^{a}u\big)^*-A^a u\big]_{i0k}
 \frac{\partial x^{\bar 2}}{\partial x^a}\Big|_{i0k}
 \delta_{j0}.
\label{eqq:411}
\end{align}
\subsection{Non-conservative equations --- weak form}
To get the weak form for the non-conservative equations,
we have to evaluate the volume term on the right-hand side of
\ref{eqq:4}):
\begin{align}
-\int u\partial_a (A^a \phi_i\sqrt{\gamma})\,d^3x&=
-\sum_{lmn,pqr}\int u_{pqr}
  \phi_{pqr}(\xbf)\left[\partial_{a}\left(A^a\sqrt{\gamma}\phi_{ijk}
  \right)_{lmn}\phi_{lmn}(\xbf)\right]\,d^3 x\nonumber\\
&=
-\sum_{pqr} u_{pqr}
  A^a_{ijk}\sqrt{\gamma_{ijk}}
  \int \phi_{pqr}(\xbf)\partial_a\phi_{ijk}(\xbf)\,d^3 x,
\end{align}
where we have used $\phi_{ijk}(\xbf_{lmn})=\delta_{il}\delta_{jm}\delta_{kn}$.
Proceeding now as for getting Eqs.\ (\ref{eq:weak3}) and (\ref{eq:weak5}), we
find that the second term on the left-hand side of
Eq.\ (\ref{eqq:411}) is replaced by
\begin{equation}
-\frac{A^a_{ijk}}{J_{ijk}}
\Big[
\sum_l \tilde D_{il}^{\bar 1}
\frac{\partial x^{\bar 1}}{\partial x^a}\Big|_{ljk}J_{ljk}
u_{ljk}
+
\sum_m \tilde D_{jm}^{\bar 2}
\frac{\partial x^{\bar 2}}{\partial x^a}\Big|_{imk}J_{imk}
u_{imk}
+
\sum_n \tilde D_{kn}^{\bar 3}
\frac{\partial x^{\bar 3}}{\partial x^a}\Big|_{ijn}J_{ijn}
u_{ijn}
\Big],
\label{eq:nonconsweak}
\end{equation}
where $\tilde D_{il}^{\bar 1}$ was defined in (\ref{eq:weak_deriv}).
In the boundary flux, the term $(A^a u)^*-A^a u$ becomes just
$(A^a u)^*$.

\subsection{Non-conservative equations --- alternative strong form}
Using the identity (\ref{eq:derivmat}), we can convert
Eq.\ (\ref{eq:nonconsweak}) to the equivalent strong form simply
by replacing $\tilde D_{il}^{\bar a}$ by $D_{il}^{\bar a}$ and changing
the sign:
\begin{equation}
\frac{A^a_{ijk}}{J_{ijk}}
\Big[
\sum_l D_{il}^{\bar 1}
\frac{\partial x^{\bar 1}}{\partial x^a}\Big|_{ljk}J_{ljk}
u_{ljk}
+
\sum_m D_{jm}^{\bar 2}
\frac{\partial x^{\bar 2}}{\partial x^a}\Big|_{imk}J_{imk}
u_{imk}
+
\sum_n D_{kn}^{\bar 3}
\frac{\partial x^{\bar 3}}{\partial x^a}\Big|_{ijn}J_{ijn}
u_{ijn}
\Big].
\label{eq:nonconsalt}
\end{equation}
When using this form,
the boundary flux is the same as in Eq.\ (\ref{eqq:411}).

We see that the alternative strong form (\ref{eq:nonconsalt}) is different
from the form given in  (\ref{eqq:411}). So as in the conservative case,
there are two inequivalent discrete formulations for the non-conservative case.

Note that in the non-conservative case, we learn nothing new by transforming
first and then integrating. The reason is that $A^{\bar a}\partial_{\bar a}
= A^{a}\partial_{a}$, without any Jacobian matrices acted on by
the differentiation operator.

\section{Numerical fluxes}
There are some issues that need to be clarified
in computing numerical fluxes when
one is working in curved spacetime (or even just in curvilinear
coordinates). We will illustrate the issues with a smooth problem,
but our conclusions also apply for fluxes that can handle discontinuities.

At a boundary, let $u^-$ denote the values of the solution in the
element itself, and let $u^+$ denote the boundary values in the neighboring
element. The aim is to construct the numerical flux using $u^-$ and $u^+$.
A key ingredient in such a recipe is the characteristic
decomposition of the variables. For Eq.\ (\ref{eq:cons4}), in regions
away from discontinuities, we can rewrite the equation in the non-conservative
form (\ref{eqq:1}) by defining
\begin{equation}
A^a = \frac{\partial F^a}{\partial u}.
\label{eq:f1}
\end{equation}
To do this, we absorb the derivatives of $\sqrt{\gamma}$ into a redefined
source term $s$. For a system of equations, $A^a$ will be a square  matrix.
To find the characteristic decomposition with respect to some normal
vector $n_a$, we diagonalize
\begin{equation}
A^a n_a=S\Lambda S^{-1}.
\label{eq:11b}
\end{equation}
Here $\Lambda$ is a diagonal matrix of the eigenvalues of $A^a n_a$,
the characteristic speeds,
and $S$ is a matrix whose columns are the corresponding eigenvectors.
Then the characteristic variables are given by the transformation $S^{-1}u$.
(The existence of this decomposition is essentially the definition of
hyperbolicity.)

Recall that for smooth problems the numerical flux is generally evaluated by
upwinding.
This means that we write $\Lambda=\Lambda^+ + \Lambda^-$, where
$\Lambda^+$ contains the positive elements of $\Lambda$ and
$\Lambda^-$ the negative elements. 
The positive speeds correspond to
variables propagating in the direction of $n_a$, that is, leaving
the element. So they are associated with characteristic
variables at the boundary
but inside the element, that is, $S^{-1}u^{-}$.
Conversely, the negative entries
$\Lambda^-$ correspond to variables propagating in the direction ${}-n_a$,
that is, entering the element. So they are associated with variables
outside the element, $S^{-1}u^{+}$.
Thus we write
\begin{equation}
(n_a A^a u)^*=(S\Lambda S^{-1}u)^*
=(S(\Lambda^+ + \Lambda^-) S^{-1}u)^*
\to S(\Lambda^+ S^{-1}u^{-}+
\Lambda^- S^{-1}u^{+}).
\label{eq:11c}
\end{equation}
Equation (\ref{eq:11c}) is the upwinding numerical flux.

Now the question arises: Which normal vector should we use in
Eq.\ (\ref{eq:11b})? Obviously, from the derivation of the DG algorithm
where we used Gauss's Theorem to get the boundary flux term, the
direction of the normal should be that of the normal to the reference
element. But what about the normalization? The standard recipe is to
use the unit normal, which in a general metric means $\gamma^{ab}n_a n_b=1$.
We will see that this leads to unnecessary complications.

Note that changing the normalization of $n_a$ in Eq.\ (\ref{eq:11b}) will
scale the characteristic speeds in $\Lambda$ by the same factor. The
normalization of the eigenvectors, by contrast, is arbitrary, since
$S$ always appears together with $S^{-1}$ in a flux prescription.
Using the unit normal follows from Gauss's Theorem and leads to the
flux prescription in Eq.\ (\ref{eq:411}), with geometric factors
like $\sqrt{{}^{(2)}\gamma}$, $\sqrt{\gamma}$,
and $J$ appearing. However, as shown
in Eq.\ (\ref{eq:412}), using the unnormalized normal vector
$n_a=\partial x^{\bar a}/\partial x^a$ removes \emph{all} the
extra geometric factors. Basically, the same factor that normalizes the
normal vector gets undone in the flux prescription.
The reason for this is explained in \ref{app:gauss}.

The reader may be concerned that many characteristic decompositions
involve \emph{products} of normal vectors. These will not scale linearly
with a rescaling of the normal vector. How can everything work out correctly?
The answer is surprising.
We illustrate with the example of a scalar wave propagating in a given
background metric.

\subsection{Scalar wave evolution} 
\label{subsec:scalar}
The scalar wave equation is
\begin{equation}
\Box \psi =\frac{1}{\sqrt{-g}}\partial_\mu\left(\sqrt{-g}g^{\mu\nu}\partial_\nu
\psi\right)=0.
\label{eq:s1}
\end{equation}
To write this in first-order form, define variables $\Phi_a$ and $\Pi$ to
replace the spatial and time derivatives
of $\psi$ as follows:
\begin{equation}
\Phi_a= \partial_a\psi\qquad
\Pi=\frac{1}{\alpha}(-\partial_t\psi+\beta^a\Phi_a).
\label{eq:s2}
\end{equation}
Then Eq.\ (\ref{eq:s1}) can be written in the form (\ref{eq:cons4}) with
\begin{equation}
u=\begin{bmatrix}
\psi\\
\Pi\\
\Phi_b
\end{bmatrix}
\qquad
F^a=\begin{bmatrix}
0\\
\alpha\gamma^{ab}\Phi_b-\Pi\beta^a\\
\alpha\Pi \delta^a{}_b-\beta^a\Phi_b
\end{bmatrix}.
\label{eq:s3}
\end{equation}
We have not bothered to write down the source terms $s$ since they are not
important for determining the numerical flux.

The next step is the characteristic decomposition. Using $F^a$ and $u$ from
\ref{eq:s3}), we determine the matrix $A^a$ from Eq.\ (\ref{eq:f1}). Then
for some normal vector $n_a$ (not necessarily a unit normal), we form
the quantity $A^a n_a$. The result is
\begin{equation}
A^a n_a = \begin{bmatrix}
-\beta^n & 0 & 0 & 0 & 0\\
0 & -\beta^n & \alpha n^x & \alpha n^y & \alpha n^z\\
0 & \alpha n^x & -\beta^n & 0 & 0\\
0 & \alpha n^y & 0 & -\beta^n & 0 \\
0 & \alpha n^z & 0 & 0 & -\beta^n
\end{bmatrix},
\label{eq:s4}
\end{equation}
where $\beta^n=\beta^a n_a$. Next we compute the eigenvalues and eigenvectors
of this matrix to form the matrices $\Lambda$ and $S$ of Eq.\ (\ref{eq:11b}).
We find
\begin{equation}
\Lambda=\diag(-\beta^n,-\beta^n,-\beta^n, -\alpha n-\beta^n,\alpha n-\beta^n),
\end{equation}
where $n$ is the magnitude of the normal vector:
$n^2=\gamma^{ab}n_a n_b$. (In a Euclidean metric, $n^2=n_x^2+n_y^2+n_z^2$.)
With a unit normal in flat spacetime, so that $\alpha=1$, $\beta^n=0$,
we see that the last two eigenvalues correspond to characteristic
speeds equal to $\pm 1$ (the speed of light). There are also three zero-speed
modes that acquire nonzero speeds when $\beta^n\neq 0$. The degeneracy
of these three modes means that the corresponding eigenvectors are not unique,
but this makes no difference since $S$ and $S^{-1}$ always appear together
in the flux formula.

Carrying out the computation in Eq.\ (\ref{eq:11c}) gives
\begin{equation}
(n_a A^a u)^* =
\begin{bmatrix}
-\beta^n \psi^- \\
\frac{1}{2}\big[n\alpha (\Pi^- - \Pi^+)-\beta^n(\Pi^- + \Pi^+)
+\alpha n^a(\Phi_a^-+\Phi_a^+) -\beta^n n^a(\Phi_a^--\Phi_a^+)/n\big]\\[3pt]
-\beta^n\Phi_b^-+\dfrac{n_b}{2n^2}
\Big\{(n\alpha+\beta^n)\big[n^a(\Phi_a^--\Phi_a^+)
+n\Pi^+\big]+(n\alpha-\beta^n)n\Pi^-\Big\}
\end{bmatrix}.
\label{eq:num_flux}
\end{equation}
Here for definiteness we have assumed $\beta^n \leq 0$; otherwise, we
interchange plus and minus variables for any term proportional to
$\beta^n$ . Now consider the scaling of the flux
terms with the normalization of $n_a$. For example, the last term in the
middle line is proportional to $\beta^n n^a/n=\beta^b n_b n^a/n$.
The factor $n^a/n$ is the
unit normal vector, so the overall scaling is with only a single power of
the scale of the normal vector.
This is true for every term in the flux. \emph{Even though we used an
unnormalized normal vector, it gets automatically normalized ``inside'' the
flux.} Thus the answer to the question of what normal vector to use
in the characteristic decomposition is that it doesn't matter as long
as its direction is correct. It is fine to use a unit normal if you want
to. However, the DG algorithm simplifies if one uses the ``external''
normal vector that gives the overall scaling of the flux to be
the quantity $n_a=\partial x^{\bar a}/\partial x^a$ for the surface
$x^{\bar a}={}$constant.

Although we have demonstrated the automatic normalization of ``internal''
normal vectors for the scalar wave equation, it seems to be a general
property of hyperbolic systems, and least in all the cases we have examined.

\section{Moving grids: Dual frames and ALE}
Many applications, both in non-relativistic terrestrial fluids and in
relativistic astrophysics, can take advantage of a moving
computational mesh. Examples include those with moving boundaries
or interfaces, but even with fixed boundaries one can expect better
accuracy if the grid absorbs some of the fluid motion. In the
case of non-relativistic conservation equations, the corresponding
methods are called ALE methods (Arbitrary Lagrangian-Eulerian), which are
extremely popular (see, e.g., \cite{donea2004} for a review).
For evolutions of Einstein's equations
with spectral methods, a moving frame was found to be necessary
\cite{scheel2006}.
In this case, the reason is that the interiors of black holes
contain singularities. However, since black hole interiors
are causally disconnected from the exterior, they can be
excised from the computational domain. Since spectral
methods use all the grid points in a subdomain, such excision
causes a problem if the black holes are moving and causing the excision
boundary to move through the domain.  Accordingly,
one uses a time-dependent map between the ``inertial frame'' in
which the black holes move and the computational grid in which the holes
remain fixed. This ``dual-frame'' method has proved very successful
(see, e.g., \cite{mroue2013}).

For DG methods, we expect to require dual frames for the same reasons that
we require them for spectral methods. In this section, we show that
dual frames can be implemented in relativity exactly as ALE is implemented
in the non-relativistic case. In fact, we show that the usual ALE
algorithm can be derived in a few lines using 4-d vector transformations.
This is in contrast to the standard derivations, e.g.,
Ref.\ \cite{thompson1985}
of 1985, where the result is equation number 121 of the derivation.
Even a streamlined
modern derivation \cite{kopriva2011} still takes 2 full journal pages.
Besides simplicity, 
another reason not to follow the standard ALE derivation here is that
it starts with the Euclidean metric of flat space, not
a general curved metric.

The derivation starts with the conservation equation (\ref{eq:cons})
written in the
form (\ref{eq:cons2}) (everything in this section goes through if we include
a source term $s$ on the right-hand side):
\begin{equation}
\frac{1}{\sqrt{-g}}\partial_t(\sqrt{-g}F^0)
+
\frac{1}{\sqrt{-g}}\partial_a(\sqrt{-g}F^a) = 0.
\label{eq:bar}
\end{equation}
Here we are in the ``inertial'' or ``physical'' frame $x^{\mu}$.\footnote{In
relativistic applications, the frame is typically inertial only at infinity.
Similarly, the coordinate system typically does not correspond directly
to physical measurements made in a local inertial reference frame except
at infinity.}

Now make a time-dependent spatial coordinate transformation to the grid frame:
\begin{equation}
\begin{split}
 t &= \hat t\\
x^{ a} &=x^{ a}(x^{\hat a},\hat t).
\end{split}
\label{eq:mapping}
\end{equation}
We use the carets for quantities in the grid frame to distinguish
the transformation from the time-independent transformation to
the reference frame used previously.
(The caret does not imply that the grid frame is orthonormal.)
It is convenient to distinguish between $t$ and $\hat t$ even though
their values are identical. When taking partial derivatives, those with
respect to $t$ keep $x^{ a}$ constant, whereas those with
respect to $\hat t$ keep $x^{\hat a}$ constant.
The Jacobian matrix of the transformation is
\begin{equation}
\frac{\partial x^{\mu}}{\partial x^{\hat\mu}}=
\begin{bmatrix}
\dfrac{\partial  t}{\partial {\hat t}} &
  \dfrac{\partial  t}{\partial x^{\hat a}}\\[9pt]
\dfrac{\partial x^{ a}}{\partial {\hat t}} &
  \dfrac{\partial x^{ a}}{\partial x^{\hat a}}
\end{bmatrix}
=
\begin{bmatrix}
\vphantom{\dfrac{\partial  t}{\partial {\hat t}}} 1 & 0 \\[9pt]
\vphantom{\dfrac{\partial x^{ a}}{\partial {\hat t}}}
v_g^{ a} &
  \dfrac{\partial x^{ a}}{\partial x^{\hat a}}
\end{bmatrix},
\label{eq:partition}
\end{equation}
where we have defined the grid velocity components as
\begin{equation}
v_g^{ a}=\dfrac{\partial x^{ a}}{\partial {\hat t}}.
\label{eq:gridveldefn}
\end{equation}
The inverse transformation is
\begin{equation}
\frac{\partial x^{\hat\mu}}{\partial x^{\mu}}
=
\begin{bmatrix}
\dfrac{\partial \hat t}{\partial {t}} &
  \dfrac{\partial \hat t}{\partial x^{a}}\\[9pt]
\dfrac{\partial x^{\hat a}}{\partial  t} &
  \dfrac{\partial x^{\hat a}}{\partial x^{ a}}
\end{bmatrix}
=
\begin{bmatrix}
\vphantom{\dfrac{\partial \hat t}{\partial {t}}} 1 &
\vphantom{\dfrac{\partial \hat t}{\partial x^{a}}} 0\\[9pt]
 -v_g^{\hat a} &
\Big(\dfrac{\partial x^{ a}}{\partial x^{\hat a}}\Big)^{-1}
\end{bmatrix}.
\label{eq:inversej}
\end{equation}
To get the second equality in (\ref{eq:inversej}) from the first, we used
the formula for the inverse of a partitioned matrix like (\ref{eq:partition})
(see, e.g., \S2.7.4 of Ref.\ \cite{numericalrecipes}).
Here we have defined the components of the grid velocity in the
grid frame just by the spatial part of the transformation:
\begin{equation}
\dfrac{\partial x^{\hat a}}{\partial  t}=-
\dfrac{\partial x^{\hat a}}{\partial x^{ a}}v_g^{ a} \equiv -v_g^{\hat a}.
\label{eq:gridvel}
\end{equation}
The first equality in (\ref{eq:gridvel})
is from the formula for the partitioned matrix inverse.

Now consider the conservation equation (\ref{eq:bar}) in the grid frame.
Since the equation is covariant, that is $\nabla_\mu F^\mu=\nabla_{\hat \mu}
F^{\hat \mu}$, we get immediately
\begin{equation}
\frac{1}{\sqrt{-\hat g}}\partial_{\hat t}(\sqrt{-\hat g}F^{\hat 0})
+
\frac{1}{\sqrt{-\hat g}}\partial_{\hat a}(\sqrt{-\hat g}F^{\hat a}) = 0.
\label{eq:grid}
\end{equation}
Here the quantity $\sqrt{-\hat g}$ can be calculated from the 4-d
equivalent of (\ref{eq:209}): $\sqrt{-\hat g}=J \sqrt{- g}$ where
from Eq.\ (\ref{eq:partition}) $J =
\det\big(\partial x^a/\partial x^{\hat a}\big)$.

To proceed,
replace the grid components $(F^{\hat 0},F^{\hat a})$ by their inertial
components:
\begin{align}
F^{\hat 0}&=\frac{\partial x^{\hat 0}}{\partial x^{\mu}} F^{\mu}
 = F^{ 0}\\
F^{\hat a}&=\frac{\partial x^{\hat a}}{\partial x^{\mu}} F^{\mu}
 = \frac{\partial x^{\hat a}}{\partial  t} F^{ 0} +
\frac{\partial x^{\hat a}}{\partial x^{ b}} F^{ b}
 = -v_g^{\hat a} F^{ 0} + \frac{\partial x^{\hat a}}{\partial x^{ b}} F^{ b}.
\label{eq:trans}
\end{align}
So eqn.\ (\ref{eq:grid}) gives the final form
\begin{equation}
\frac{1}{\sqrt{-\hat g}}\partial_{\hat t}(\sqrt{-\hat g}F^{0})
+
\frac{1}{\sqrt{-\hat g}}\partial_{\hat a}\left[\sqrt{-\hat g}
\left(\frac{\partial x^{\hat a}}{\partial x^{ b}} F^{ b}
-v_g^{\hat a} F^{ 0}\right)\right]
 = 0.
\label{eq:final}
\end{equation}
This equation is again in flux-conservative form. We can use it with the vector
components in the inertial frame as written, or in the grid frame.
In that case, $F^{ 0}=F^{{\hat 0}}$,
but we can regard $F^{\hat a}$ as being a purely spatial vector given by
\begin{equation}
F^{\hat a}=\frac{\partial x^{\hat a}}{\partial x^{ b}} F^{ b}\quad\text{(spatial
transformation only),}
\end{equation}
rather than as a piece of a 4-vector with transformation given by
eqn.\ (\ref{eq:trans}).

We can use Eq.\ (\ref{eq:final}) in the DG algorithm in several ways.
One way is to introduce another time-independent map from the grid frame
(hatted) to the reference frame (barred) and follow the previous
treatment. Alternatively, we can choose the mapping to the grid frame
to go all the way to the reference frame and work with the equation
directly in that frame (``transform-then-integrate,''
cf.\ \S\ref{sec:transform}).

While keeping conservation form with a moving mesh is important
for equations that are fundamentally conservation laws, for non-conservative
systems like (\ref{eqq:1}) one simply leaves the extra terms outside
the derivative operators since there is no simple way to remove their
effects completely.

\subsection{ALE, or the ``wrong'' way to derive the equations}
In non-relativistic fluid dynamics, mappings like
(\ref{eq:mapping}) define ALE methods.
As mentioned earlier, the derivations
of the ALE equations in the literature are quite complicated
\cite{thompson1985,kopriva2011}, with many
papers just referring back to the original derivations.
These derivations start with eqn.\ (\ref{eq:bar}) with
$ g=1$ because the coordinates are Cartesian.
Using the chain rule, the equation becomes
\begin{align}
\frac{\partial F^{ 0}}{\partial  t}
+\frac{\partial F^{ a}}{\partial x^{ a}} &=
  \frac{\partial F^{ 0}}{\partial {\hat t}}
+ \frac{\partial x^{\hat a}}{\partial  t}\frac{\partial F^{ 0}}{\partial x^{\hat a}}
+\frac{\partial x^{\hat a}}{\partial x^{ a}}
  \frac{\partial F^{ a}}{\partial x^{\hat a}}\nonumber\\
&= \frac{\partial F^{ 0}}{\partial {\hat t}} -v_g^{\hat a}
\frac{\partial F^{ 0}}{\partial x^{\hat a}}
+\frac{\partial x^{\hat a}}{\partial x^{ a}}
  \frac{\partial F^{ a}}{\partial x^{\hat a}}.
\end{align}
To get an equation in conservation form, one rewrites the second and third
terms with everything inside the $\partial/\partial x^{\hat a}$ minus some extra
terms. Then one shows that the extra terms can all be absorbed by putting
$\sqrt{\hat g}=J$ inside the
$\partial/\partial {\hat t}$. The manipulations
involve complicated maneuvers with derivatives of the Jacobian matrix.
The final result is exactly eqn.\ (\ref{eq:final}).
In this approach, the fact that the final result is again in
conservative form is a surprise.
The miraculous cancellations that occur involve the metric identities
\ref{eq:504})
and the \emph{geometric conservation law} (see \ref{app:geom}).

The covariant derivation, by contrast, just uses straightforward properties of
the divergence operator and the fact that $ t = {\hat t}$. It does not assume
that the 4-d metric is that of relativity. All we need to assume is that the
fundamental equation is a 4-divergence with respect to some metric. Then,
since $ t = \hat t$, the final equation involves
tensor transformations that are
purely spatial, using only the spatial part of the Jacobian matrix.
In Newtonian physics, physical quantities are required to be
tensors only under such spatial transformations,
so the final result is valid, with no mystery as to why the transformed
equation should be a conservation law.

\section{Numerical experiments}
\label{sec:expt}
\begin{figure}[ht]
\centering
\includegraphics[width=2in]{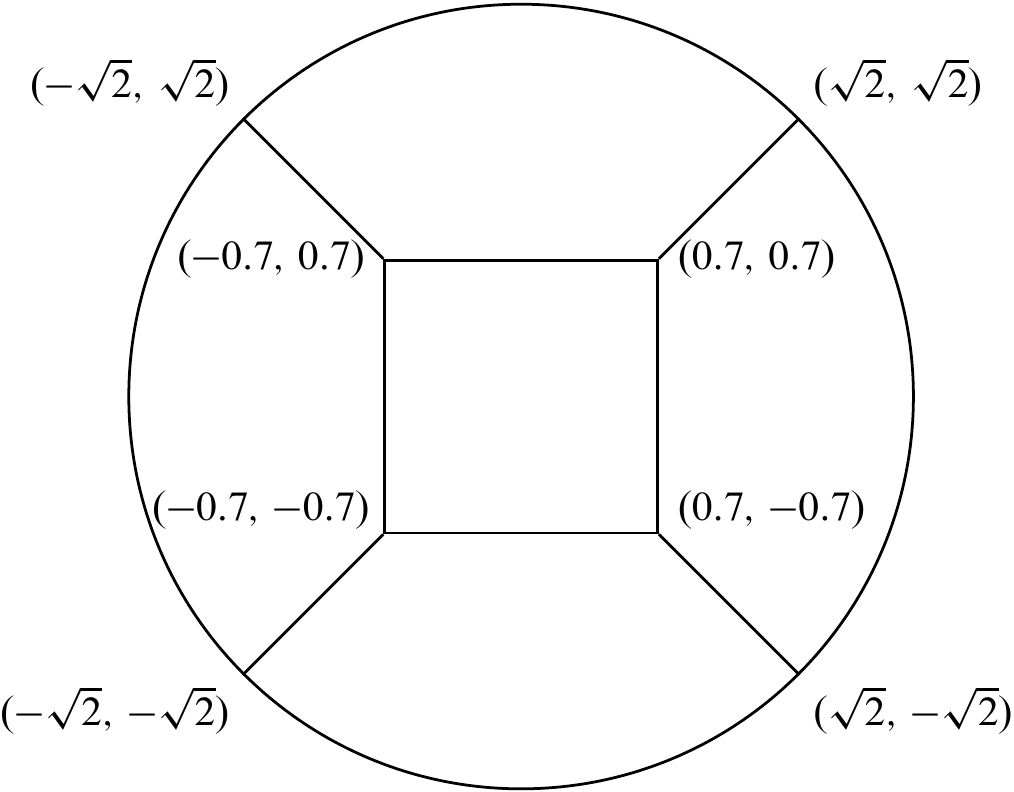}
\caption{Domain of 5 elements used for numerical experiments.}
\label{fig:disk}
\end{figure}

The algorithms developed above are intended to be used for fully
relativistic 3-dimensional problems. Here, we investigate them for
a simple test problem to highlight some of their properties.
We have
implemented the algorithms for the scalar wave equation of \S\ref{subsec:scalar}
in flat spacetime, that is, with $\alpha=1$, $\beta=0$, and $\gamma_{ab}=
\delta_{ab}$. The domain is a 2-dimensional disk divided into five elements 
as shown in Figure~\ref{fig:disk}.
Such a domain was used in Chapter 8 of Ref.\ \cite{kopriva2009} for a
sightly different test problem. We study two different ways of
mapping the elements with curved boundaries
to the reference square. First, following Chapter 6 of
Ref.\ \cite{kopriva2009},
we use an isoparametric map with transfinite blending.
In this case, the map is constructed with polynomials of degree $N$
and since we are in two dimensions,
the discrete version of the metric identities 
holds identically without any special precautions \cite{kopriva2006}.

The second map is analytic and depends on four parameters.
The quantities  $e_{\rm min}$
and $e_{\rm max}$ give the maximum values of the $x$ locations of the
inner and outer boundaries. For the rightmost element in
Figure~\ref{fig:disk}, these are 0.7 and 2.
The quantities $c_{\rm min}$
and $c_{\rm max}$ describe the curvature of the inner and outer boundaries
of the wedge, respectively. If $c_{\rm min}=0$, the inner edge
is flat, whereas if $c_{\rm min}=1$, it is
circular. Similarly, if $c_{\rm max}=0$, the outer edge is flat, whereas
$c_{\rm max}=1$ corresponds to a circular outer edge.
Using these parameters, we define
\begin{equation}
\begin{split}
x_{\rm min} &= \frac{e_{\rm min}c_{\rm min}}{\sqrt{1+\bar y^2}}
  + (1-c_{\rm min})e_{\rm min}\\
x_{\rm max} &= \frac{e_{\rm max}c_{\rm max}}{\sqrt{1+\bar y^2}}
  + (1-c_{\rm max})e_{\rm max}.
\end{split}
\end{equation}
Then the map is given by
\begin{equation}
\begin{split}
x&=x_{\rm min}+(x_{\rm max}-x_{\rm min})\frac{\bar x-e_{\rm min}}{e_{\rm max}
-e_{\rm min}}\\
y&=x\bar y.
\end{split}
\end{equation}
Note that in these expressions $\bar x \in [e_{\rm min},e_{\rm max}]$.
A simple linear map puts $\bar x$ in the standard range $[-1,1]$.
Permutations of $x$ and $y$ and sign changes allow all four curvilinear
elements to be generated with the above map.

For one set of experiments with this second map,
we compute the Jacobian of the transformation analytically.
In this case, the discrete metric identities are not satisfied. To
see how important these identities are,  in a second set of
experiments we evaluate the Jacobian by applying the differentiation matrix
(\ref{eq:405}) directly to the $x$ and $y$ values at the collocation points.
Since this corresponds to differentiating the interpolating polynomial of
degree $N$, the result is also a polynomial and in two dimensions
the discrete metric identities are satisfied \cite{kopriva2006}. 

The test case is a wave propagating at $45^\circ$ to the coordinate
axes, with analytic solution
\begin{equation}
\psi=A\sin(t-\kbf\cdot\xbf).
\end{equation}
Here $A=1$ and $k_x=k_y=1/\sqrt{2}$. The numerical flux is computed
using Eq.\ (\ref{eq:num_flux}). Where the boundary conditions require
solution values outside the computational
domain, these are provided by the analytic solution.
We integrate from $t=0$ to $t=1$ with a fixed timestep $dt=2\times
10^{-4}$ using a
low-storage third-order Runge-Kutta method (Case 7 in \cite{williamson1980}).
All elements have the same spatial
resolution, which we vary from $N=4$ to $N=20$.
For these choices, the timestepping error is negligible compared with the
error from the spatial discretization.

\begin{figure}[ht]
\centering
\includegraphics[width=3in,angle=-90]{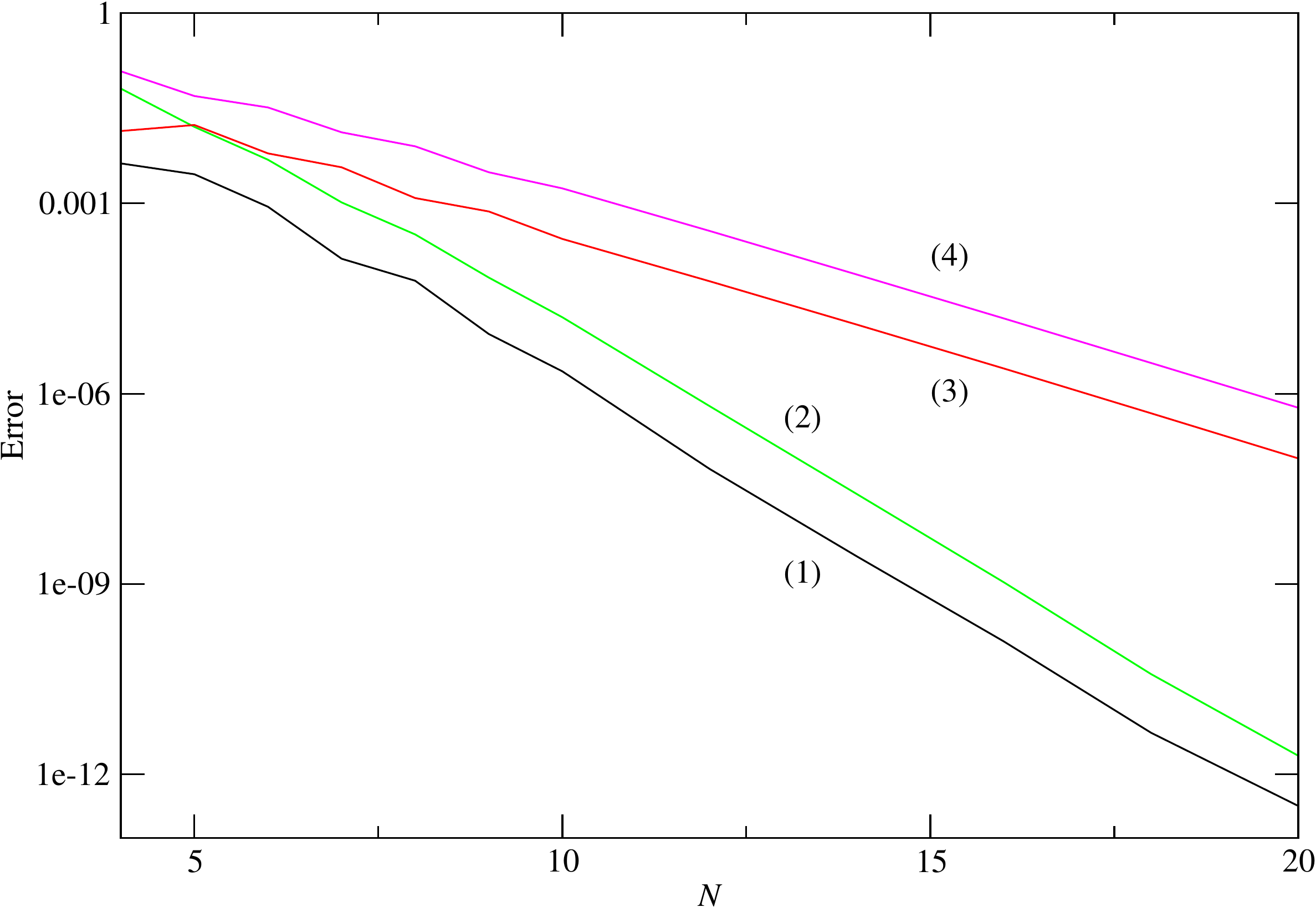}
\caption{Results of numerical experiments for different DG implementations.
Shown is the error as a function of resolution $N$ for integrating the
scalar wave equation as described in the text.
The numerical annotations on each line correspond to cases (1) -- (4)
in the text.
}
\label{fig:results}
\end{figure}

So altogether there are 6 experiments:
\begin{enumerate}
\item
Integrate first (Eqs.\ \ref{eq:411} and \ref{eq:bdy}) with isoparametric map.
\item
Transform first (Eq.\ \ref{eq:509}) with isoparametric map.
\item
Integrate first with analytic map and analytic Jacobian.
\item
Transform first with analytic map and analytic Jacobian.
\item
Integrate first with analytic map and numerical Jacobian satisfying metric
identities.
\item
Transform first with analytic map and numerical Jacobian satisfying metric
identities.
\end{enumerate}
The results are shown in Figure~\ref{fig:results}.

We find that for the analytic map used here, the results when the metric
identities are satisfied are indistinguishable on the plot from
the results with the analytic Jacobian. Accordingly, experiments (5) and (6)
are not shown in the figure. For the analytic map, it appears that the
grid distortion is small enough that the metric identity error is
less than the truncation error. We explicitly computed the discrete
metric identities for these cases to verify that, although they do not
vanish, their magnitude is smaller than the truncation error.

The first thing we note from Figure~\ref{fig:results} is the expected
exponential convergence of the error with resolution for all the
implementations. Cases (1) and (2), which use the isoparametric map,
have smaller errors for the same resolution than the corresponding
cases (3) and (4) that use the analytic map. As explained in the previous
paragraph, this is \emph{not} because the isoparametric map satisfies
the metric identities. Rather, the isoparametric map introduces less
grid distortion for this example. We can quantify this very roughly
by examining the magnitude of the Jacobian, which is closer to unity
by about a factor of 2 for the isoparametric map than the analytic map.

The most interesting result from these experiments is that, independent
of the mapping, the integrate-then-transform formulation performs
somewhat better than the transform-then-integrate version.
The error for a given $N$ is consistently smaller for cases (1) and (3)
compared with (2) and (4). It should be noted, however,
that for this simple
example, the computation time is dominated by the matrix-vector multiplies
in computing the derivatives. For cases (1) and (3), each derivative
matrix has to operate on each component of the flux vector, whereas
for cases (2) and (4) each derivative matrix operates on only a single
component of the flux vector. Accordingly, the
computation time is twice as long (in two dimensions) for cases
(1) and (3) than for cases (2)
and (4). Realistic astrophysics applications are likely to be dominated
by complicated equation of state calculations or source terms, so
it is not clear that this property will be important in practice.
Since the transform-then-integrate algorithm is widely used,
whether integrate-then-transform is better just for this setup or whether it is
better in general needs to be examined in future work.

Note that for this example, integrating the scalar wave equation in
non-conservative form gives identical results to the conservative form
studied in the experiments. This is because for linear equations
with constant coefficients the
matrix $A^a = \partial F^a/\partial u$ is constant and so the two
formulations are essentially identical. 

We have not investigated the effect of enforcing the
geometric conservation law in the case where one uses a
time-dependent map. There are various strategies  for doing so.
These strategies, and the accuracy of the resulting algorithms, are
discussed for example in Ref.\ \cite{kopriva2011}.

\section{Conclusions}
We have shown that because of the underlying covariance of the
equations, the DG algorithm can easily be formulated in general
relativity.
The formulation turns out to be very similar to
that for curvilinear coordinates in flat spacetime, which leads
to several insights that are applicable in that arena as well.

We find that in general there
are two distinct strong formulations for conservation laws.
These correspond to integrate-then-transform, Eqs.\ (\ref{eq:411}) and
(\ref{eq:bdy}), and transform-then-integrate, Eq.\ (\ref{eq:509}).
Both weak forms are equivalent to this second strong form
for the tensor-product basis functions we use. In one spatial dimension,
or for the common case of a Cartesian grid with elements
simply translated in $x$, $y$, and $z$, the two strong formulations
are equivalent.

Most applications in curvilinear coordinates in flat space have used
the formulation (\ref{eq:509}). However, in the numerical experiments
described in \S\ref{sec:expt}, the other formulation
was more efficient. Whether this is true or not in general remains
to be seen.

For hyperbolic equations in non-conservation form, there are also
two inequivalent formulations, Eqs.\ (\ref{eqq:411})
and (\ref{eq:nonconsalt}). These formulations are important
for solving Einstein's equations, which generally are not written in
conservation form.

We have given a careful discussion of how numerical fluxes should
be handled. In particular, we have shown that the normal
vector that the boundary flux vector is projected along does not
need to be the unit normal---the normalization factor cancels
out of the algorithm. We call this normal vector the ``outside'' normal
vector. If a characteristic decomposition is used
to construct the numerical flux vector, it is the \emph{unit}
normal that appears
``inside'' the flux. However, it is not necessary to explicitly normalize
this vector either: Any normal vector that is used in finding
the characteristic decomposition gets automatically normalized ``inside''
the flux.

We have shown that moving grids implemented with
time-dependent mappings (ALE  and dual-frame methods)
are easily handled in the relativistic treatment.
We give a novel derivation of the ALE method
that uses general covariance to get the result in a few lines.
In addition,
the reason that the ALE method preserves the conservation form of the
equations is explained.

We clarify several aspects of the metric identities. For example,
we show that
satisfying the metric identities discretely is a
necessary condition for free-stream preservation
only for one of the
computational formulations of the DG algorithm and not the other.
The numerical experiments in \S\ref{sec:expt} suggest that
satisfying the metric identities is not necessarily a prerequisite for
accurate simulations, but likely depends on the problem.

The formulation of the DG method in this paper will allow the algorithm
to be applied to general problems in computational astrophysics that
involve relativistic gravity. These methods hold great promise
for achieving high accuracy and efficiency on current
and upcoming supercomputer
hardware. It will be interesting to see how well they perform in practice.

\section*{Acknowledgments}
I thank Fran\c{c}ois H\'ebert for helpful discussions.
I am grateful for the hospitality of TAPIR and the Walter Burke Institute for
Theoretical Physics at Caltech where part of this work was carried out.
This work was supported in part by
NSF Grants PHY-1306125 and AST-1333129 at Cornell University, and
by a grant from the Sherman Fairchild Foundation.

\appendix
\makeatletter
\renewcommand\thesubsection{\@Alph\c@section.\@arabic\c@subsection}
\makeatother
\section{Gauss's Theorem and normal vectors}
\label{app:gauss}

The usual form of Gauss's Theorem for an arbitrary curved (or curvilinear)
metric is
\begin{equation}
\int_V \nabla_a F^a\sqrt{\gamma}\,d^3x =
\oint_S F^a \,d^2\Sigma_a
=\oint_S F^a n_a\,d^2\Sigma.
\label{eq:214}
\end{equation}
This is a physically appealing result: The left-hand side is the divergence
of a flux integrated over proper volume. The right-hand side is the total
flux through the surface. In this form, the result is manifestly independent
of the coordinates. Let's now spoil this elegant result.

The covariant divergence in the integrand of Eq.\ (\ref{eq:214})
can be written as
\begin{equation}
\nabla_a F^a=\frac{1}{\sqrt{\gamma}}\partial_a\left(\sqrt{\gamma}F^a\right).
\label{eq:213}
\end{equation}
So, letting $\sqrt{\gamma}F^a \to F^a$, an equivalent form of (\ref{eq:214})
is
\begin{equation}
\int_V \partial_a F^a\,d^3x =
\oint_S F^a \frac{1}{\sqrt{\gamma}}\,d^2\Sigma_a
=\oint_S F^a n_a\frac{1}{\sqrt{\gamma}}\,d^2\Sigma.
\label{eq:201}
\end{equation}
We see explicitly that Gauss's Theorem appears to depend
on the metric in several places, for example,
to define the unit normal $n_a$ and the invariant surface element $d^2\Sigma$.
However, it turns out that this dependence is illusory, introduced originally
by Stokes who first developed the geometric form of the theorem.
For our application, it is actually more
convenient to revert to the primitive form of the theorem that uses only
basic calculus and does not rely on the metric.

Start with one of the terms on the left-hand side of Eq.\ (\ref{eq:201}):
\begin{align}
I_3&=\int \partial_3 F^3\,d^3x\nonumber\\
&=\int \frac{\partial F^3}{\partial x^3}
\,dx^1\,dx^2\,dx^3
\nonumber\\
&=\int F^3\Big|^{x^3_+}_{x^3_-}
\,dx^1\,dx^2.
\label{eq:200a}
\end{align}
Consider first the surface $x^3=x^3_+$.
Choose intrinsic coordinates $(a^1,a^2)$
to parametrize this surface, so that on the surface
\begin{equation}
\begin{split}
x^1&=x^1(a^1,a^2)\\
x^2&=x^2(a^1,a^2).
\end{split}
\label{eq:200b}
\end{equation}
Extend the coordinates $(a^1,a^2)$ to cover the neighborhood of the
surface by introducing
a third coordinate $a^3$ such that the boundary is a level surface
$a^3=\text{constant}$.
The contribution to the integral (\ref{eq:200a}) is
\begin{align}
I_{3+}&=\int F^3(x^3_+)\,dx^1\,dx^2\nonumber\\
&=\int F^3 \frac{\partial(x^1,x^2)}{\partial(a^1,a^2)}\,da^1\,da^2.
\label{eq:200c}
\end{align}

An alternative form can be derived using some Jacobian gymnastics:
\begin{align}
\left.\frac{\partial(x^1,x^2)}{\partial(a^1,a^2)}\right|_{a_3}
&=
\frac{\partial(x^1,x^2,a^3)}{\partial(a^1,a^2,a^3)}\nonumber\\
&=
\frac{\partial(x^1,x^2,a^3)}{\partial(x^1,x^2,x^3)}
\frac{\partial(x^1,x^2,x^3)}{\partial(a^1,a^2,a^3)}\nonumber\\
&= J\left.\frac{\partial a^3}{\partial x^3}\right|_{x_1,\, x_2}.
\label{eq:200d}
\end{align}
(You can also derive Eq.\ (\ref{eq:200d}) by noting that $\partial a^3/
\partial x^3$ is an element of the inverse of the Jacobian matrix.
Eq.\ (\ref{eq:200d}) says that this element is given by the cofactor
of $\partial x^3/ \partial a^3$
divided by the determinant of the matrix.)

Substituting Eq.\ (\ref{eq:200d}) in Eq.\ (\ref{eq:200c}) gives
\begin{equation}
I_{3+}=\int F^3 \frac{\partial a^3}{\partial x^3} J\, da^1\,da^2.
\end{equation}
Summing up contributions like this from each segment of the boundary surface
gives the simple form of Gauss's Theorem that is in effect
being used in the main text, with the identification $J\, da^1\,da^2 \to
J\, dx^{\bar 1}\,dx^{\bar 2} = dx^1\,dx^2$.

\subsection{Using a metric}
The form (\ref{eq:200c}) or (\ref{eq:200d}) does not make use of a metric.
We can start to see where the usual form of Gauss's theorem comes from
by noting that the normal to the surface $a^3=\text{constant}$ is proportional
to $(0,0,1)$ in the $(a^1,a^2,a^3)$ coordinates, and so $\partial a^3/\partial
x^3$ is proportional to the component $n_3$ in the $(x^1,x^2,x^3)$
coordinates. But to write the result in terms of a true unit normal we
need to introduce a metric.
To do this, note that
\begin{align}
\frac{\partial(x^1,x^2)}{\partial(a^1,a^2)}\,da^1\,da^2 &=
e_{3ab}\frac{\partial x^a}{\partial a^1}\frac{\partial x^b}{\partial a^2}
\,da^1\,da^2\nonumber\\
&=\frac{1}{2}e_{3ab}\,dx^a\wedge dx^b\nonumber\\
&=\frac{1}{\sqrt{\gamma}}\frac{1}{2}\epsilon_{3ab}\,dx^a\wedge dx^b.
\label{eq:202}
\end{align}
Here $e_{abc}$ is the completely antisymmetric
permutation symbol and $\epsilon_{abc}=\sqrt{\gamma}e_{abc}$
is the permutation tensor (Levi-Civita tensor).
The 1-form $dx^a$
is defined from the parametrization of the surface, $x^a=x^a(a^1, a^2)$:
\begin{equation}
dx^a=\frac{\partial x^a}{\partial a^1}da^1+
\frac{\partial x^a}{\partial a^2}da^2.
\label{eq:202a}
\end{equation}

Define the surface element in the usual way:
\begin{equation}
d^2\Sigma_c=\frac{1}{2}\epsilon_{cab}\,dx^a\wedge dx^b.
\label{eq:203}
\end{equation}
This requires only a notion of volume ($\sqrt{\gamma}$).
With a full metric, however, it is equivalent to
\begin{equation}
d^2\Sigma_c=n_c d^2\Sigma,\qquad d^2\Sigma=\sqrt{{}^{(2)}\gamma}da^1\,da^2.
\label{eq:204}
\end{equation}
as we will verify below.
The vector with components $n_c$ is the normal to the surface, here assumed
to be $a^3=\text{constant}$, and ${}^{(2)}\gamma$ is the determinant of the
2-dimensional metric induced on the surface by $\gamma_{ij}$.

\subsection{Proof that $d^2\Sigma_c=n_c d^2\Sigma$}
First get an expression for the normal vector.
Let $x^{\bar a}$ denote the coordinates $(a^1,a^2,a^3)$. The unit normal
to the surface $a^3=\text{constant}$ has components
\begin{equation}
n_{\bar a}=\frac{1}{\sqrt{\gamma^{\bar 3\bar 3}}}(0,0,1),
\label{eq:206}
\end{equation}
where
\begin{equation}
\gamma^{\bar 3\bar 3}=\frac{\text{cofactor}
(\gamma_{\bar 3\bar 3})}{\det(\gamma_{\bar i
\bar j})}=\frac{{}^{(2)}\gamma}{\bar \gamma}.
\label{eq:207}
\end{equation}
Taking determinants of both sides of the transformation rule
\begin{equation}
\gamma_{\bar i \bar j}=\frac{\partial x^i}{\partial x^{\bar i}}
\frac{\partial x^j}{\partial x^{\bar j}}\gamma_{ij}
\label{eq:208}
\end{equation}
gives the transformation rule
\begin{equation}
\sqrt{\bar \gamma}=J\sqrt{\gamma}.
\label{eq:209}
\end{equation}
Thus Eq.\ (\ref{eq:206}) becomes
\begin{equation}
n_{\bar a}=\frac{\sqrt{\bar \gamma}}{\sqrt{{}^{(2)}\gamma}}(0,0,1)=
J\frac{\sqrt{\gamma}}{\sqrt{{}^{(2)}\gamma}}(0,0,1),
\label{eq:210}
\end{equation}
and so
\begin{align}
n_a&=\frac{\partial x^{\bar a}}{\partial x^a}n_{\bar a}\nonumber\\
&=\frac{\partial a^3}{\partial x^a}J\frac{\sqrt{\gamma}}{\sqrt{{}^{(2)}\gamma}}.
\label{eq:211}
\end{align}

Now we can verify Eq.\ (\ref{eq:204}):
\begin{alignat}{2}
d^2\Sigma&=n^c d^2\Sigma_c\nonumber\\
&=\frac{1}{2}n^c \epsilon_{cab}\,dx^a\wedge dx^b\nonumber\\
&=\frac{1}{2}n^{\bar c} \epsilon_{\bar c\bar a\bar b}\,
dx^{\bar a} \wedge dx^{\bar b}&\qquad&\text{(since the quantity is a
scalar)}\nonumber\\
&= n^{\bar 3}\sqrt{\bar \gamma}\,da^1\,da^2\nonumber\\
&= \gamma^{\bar 3 \bar 3}n_{\bar 3}\sqrt{\bar \gamma}\,da^1\,da^2\nonumber\\
&=\sqrt{{}^{(2)}\gamma}\,da^1\,da^2&\qquad&\text{(using (\ref{eq:207}) and
(\ref{eq:210}))},
\label{eq:212}
\end{alignat}
which is the correct expression for the surface element.

\subsection{The usual form of Gauss's Theorem}
Now return to Eq.\ (\ref{eq:200c}).
Equation (\ref{eq:202}) becomes
\begin{equation}
\frac{\partial(x^1,x^2)}{\partial(a^1,a^2)}\,da^1\,da^2=
\frac{1}{\sqrt{\gamma}}d^2\Sigma_3=
\frac{1}{\sqrt{\gamma}}n_3\,d^2\Sigma.
\label{eq:205}
\end{equation}
So Eq.\ (\ref{eq:200c}) can be rewritten as
\begin{equation}
I_{3+}=\int F^3 n_3\frac{1}{\sqrt{\gamma}}\,d^2\Sigma.
\label{eq:200g}
\end{equation}
A similar expression holds on the lower surface $x^3=x^3_-$ provided
we interpret the normal vector as pointing in the outward direction.
This is because the outward normal has components proportional to
$(0,0,-1)$, but
$F^3(x^3_-)$ appears in the integrand of (\ref{eq:200a}) with a
compensating minus sign.

Similar arguments hold for the terms $I_1$ and $I_2$ analogous to
(\ref{eq:200a}). The conclusion is that Gauss's Theorem in the form
Eq.\ (\ref{eq:201}) holds in general.

To complete the proof of Gauss's Theorem, we would need to deal with
concave surfaces (above we implicitly assumed each boundary surface
was convex) and also surfaces with corners, where $x_+$ and $x_-$
can be on boundary segments parametrized by different pairs
of $a^1$, $a^2$, and $a^3$. Both these complications can be handled
by subdividing the volume into subvolumes, and the details are
not important for our purposes.

\section{The metric identities}
\label{app:identities}
The metric identities (\ref{eq:504}) follow from properties of the Jacobian
matrix itself. First,
an element of the inverse Jacobian matrix $\partial  x^{\bar a}
/\partial x^a$ is given by the cofactor of $\partial  x^a
/\partial x^{\bar a}$ divided by the determinant $J$.
We can recover this result using some Jacobian manipulations:
If $a$, $b$, and $c$ are in cyclic order, then
\begin{align}
\left.\frac{\partial x^{\bar a}}{\partial x^a}\right|_{x^b,\, x^c}
&=
\frac{\partial(x^{\bar a},x^b ,x^c)}{\partial(x^a,x^b,x^c)}
\notag\\
&=
\frac{\partial(x^{\bar a},x^b ,x^c)}{\partial(x^{\bar a},x^{\bar b},x^{\bar c})}
\frac{\partial(x^{\bar a},x^{\bar b},x^{\bar c})}{\partial(x^a,x^b,x^c)}
\notag\\
&=
\frac{\partial(x^b ,x^c)}{\partial(x^{\bar b},x^{\bar c})}\frac{1}{J}.
\end{align}
Equivalently,
\begin{equation}
J\frac{\partial x^{\bar a}}{\partial x^a}=
\epsilon_{abc}\frac{\partial x^b}{\partial x^{\bar b}}
\frac{\partial x^c}{\partial x^{\bar c}}.
\end{equation}
We can rewrite this equation to build in the cyclic order requirement:
\begin{equation}
J\frac{\partial x^{\bar a}}{\partial x^a}=
\frac{1}{2}\epsilon^{\bar a \bar b \bar c}
\epsilon_{abc}\frac{\partial x^b}{\partial x^{\bar b}}
\frac{\partial x^c}{\partial x^{\bar c}}.
\label{eq:metric2}
\end{equation}
The metric identity follows immediately from the expression (\ref{eq:metric2}).
For when we take its divergence with $\partial_{\bar a}$, the second
derivative terms are symmetric in $\bar a$ and $\bar b$ or $\bar a$ and
$\bar c$, whereas $\epsilon^{\bar a \bar b \bar c}$ is antisymmetric.

Note that in the general  derivation given here,
the elements of the Jacobian matrix
cannot be interpreted as components of basis vectors in curvilinear
coordinates. This is because we are not starting from a Euclidean metric.

\section{An identity for GLL derivative matrices}
\label{app:derivmat}
Equation (\ref{eq:derivmat}) is an identity for Gauss-Legendre-Lobatto
derivative matrices that is useful in converting between the strong
and weak forms of the DG equations,
and is also at the root of discrete ``summation by parts.''
The relation is implicit in the second equation following Eq.\ (25) in
Ref.\ \cite{carpenter1996}.
Here we demonstrate the identity explicitly.

Consider the integral on the reference element that defines the ``stiffness
matrix,''
\begin{equation}
S_{il}=\int_{-1}^1 \ell_i(x)\ell_l'(x)\,dx.
\label{eq:stiff1}
\end{equation}
Integrating by parts gives
\begin{align}
S_{il}&=\left.\ell_i(x)\ell_l(x)\right|^1_{-1}-\int_{-1}^1
  \ell_i'(x)\ell_l(x)\,dx\notag\\
&=\delta_{Ni}\delta_{Nl}-\delta_{0i}\delta_{0l}-S_{li}.
\label{eq:stiff}
\end{align}
To get the last line, note that for the Lobatto
case,  all the cardinal functions  must
vanish at $\pm1$ except for $\ell_0$ and $\ell_N$.
Now for $N+1$ grid points, the integrand in (\ref{eq:stiff1}) is a
polynomial of degree
$2N+1$, and so doing the integral by Gauss-Lobatto
quadrature gives the exact result:
\begin{align}
S_{il}&=\sum_j w_j \ell_i(x_j)\ell_l'(x_j)\notag\\
&=\sum_j w_j \delta_{ij} D_{jl}\notag\\
&=w_iD_{il}.
\label{eq:stiff2}
\end{align}
Substituting (\ref{eq:stiff2}) in (\ref{eq:stiff}) gives the relation
\ref{eq:derivmat}).

\section{The geometric conservation law}
\label{app:geom}
The geometric conservation law is an identity for the Jacobian of a
time-dependent spatial coordinate transformation like (\ref{eq:mapping}).
From Eq.\ (\ref{eq:partition}), the Jacobian is the determinant of
the spatial Jacobian matrix:
\begin{equation}
J=\det\Jbf, \qquad \Jbf=\frac{\partial x^a}{\partial x^{\hat a}}.
\end{equation}

To derive the identity, use the formula for the derivative of
a determinant,
\begin{align}
\partial_{\hat t} J &= J\Tr \left(\Jbf^{-1}\cdot\partial_{\hat t}\Jbf\right) 
\nonumber\\
&=J\left(J_{ba}^{-1}\partial_{\hat t}J_{ab}\right)\qquad
\text{(sum on $a$ and $b$)}.
\end{align}
Using Eq.\ (\ref{eq:gridveldefn}) and replacing the index $b$ by $\hat a$
gives the geometric conservation law:
\begin{align}
\frac{\partial J}{\partial\hat t}
&= J\frac{\partial x^{\hat a}}{\partial x^a}
   \frac{\partial v_g^a}{\partial x^{\hat a}}\nonumber\\
&= \frac{\partial }{\partial x^{\hat a}}\left(J \frac{\partial x^{\hat a}}
  {\partial x^a} v_g^a\right) \qquad\text{(using Eq.\ \ref{eq:504})}\nonumber\\
&=\frac{\partial }{\partial x^{\hat a}}\left(J v_g^{\hat a}\right).
\label{eq:geom_cons}
\end{align}

\section*{References}

\end{document}